\documentclass[preprint,12pt]{elsarticle}

\usepackage{amssymb}
\usepackage{amsmath}
\usepackage{amsfonts}
\usepackage{siunitx}
\usepackage{empheq}
\usepackage{bm}
\usepackage[cal=boondoxupr,frak=euler]{mathalpha}

\usepackage{graphicx}
\usepackage{newtxtext}
\usepackage{newtxmath}
\usepackage{natbib}
\usepackage{hyperref}
\hypersetup{
    colorlinks = true,
    urlcolor   = blue,
    citecolor  = black,
}
\usepackage[capitalise]{cleveref}
\usepackage{comment}
\usepackage[version=4]{mhchem}
\usepackage{lineno}
\newcommand{\appropto}{\mathrel{\vcenter{
  \offinterlineskip\halign{\hfil$##$\cr
\propto\cr\noalign{\kern2pt}\sim\cr\noalign{\kern-2pt}}}}}

\journal{Combustion and Flame}

\begin{document}

\begin{frontmatter}



\title{A quantitative theory for heterogeneous combustion of nonvolatile metal particles in the diffusion-limited regime}
%at small Reynolds numbers}


\author{Daoguan Ning\corref{cor1}}
\ead{ning@rsm.tu-darmstadt.de}
\author{Andreas Dreizler}

\address{Reactive Flows and Diagnostics, Department of Mechanical Engineering, Technical Univeristy of Darmstadt, Otto-Berndt-Str. 3, 64287 Darmstadt, Germany}

\begin{abstract}
The paper presents an analytical theory quantitatively describing the heterogeneous combustion of nonvolatile (metal) particles in the diffusion-limited regime. It is assumed that the particle is suspended in an unconfined, isobaric, quiescent gaseous mixture and the chemisorption of the oxygen takes place evenly on the particle surface. The exact solution of the particle burn time is derived from the conservation equations of the gas-phase described in a spherical coordinate \textcolor{black}{system} with the utilization of constant thermophysical properties, evaluated at a reference film layer. This solution inherently takes the Stefan flow into account. The approximate expression of the time-dependent particle temperature is solved from the conservation of the particle enthalpy by neglecting the higher order terms in the Taylor expansion of the product of the transient particle density and diameter squared. Coupling the solutions for the burn time and time-dependent particle temperature provides quantitative results when initial and boundary conditions are specified. The theory is employed to predict the burn time and temperature of micro-sized iron particles, which are then compared with measurements, as the first validation case.
The theoretical burn time agrees with the experiments almost perfectly at both low and high oxygen levels. The calculated %time-dependent 
particle temperature matches the measurements fairly well at relatively low oxygen mole fractions,  whereas the theory overpredict the particle peak temperature due to the neglect of evaporation and the possible transition of the combustion regime.
\end{abstract}


\end{frontmatter}

%\linenumbers

\section*{Novelty and significance statement}
For the first time, we present a comprehensive and quantitative analytical theory elucidating the heterogeneous combustion of nonvolatile (metal) particles in the diffusion-limited regime.
This novel theoretical model exhibits a remarkable capacity for quantitative prediction, obviating the need for supplementary information from numerical simulations or experimental data.
The derivation process of analytical solutions for burn time and time-dependent particle temperature from conservation equations is elaborated, offering transparency and insight into the model's foundations.
To demonstrate the practical utility of the theory, we apply it to analyze the combustion of iron particles, providing valuable mathematical perspectives on the underlying processes. The model's predictions for burn time and temperature align closely with experimental results, offering a partial validation of the theory within the realm of its applicable assumptions. This pioneering work contributes a robust and versatile analytical framework, advancing our understanding of diffusion-limited combustion phenomena of nonvolatile particles.

\section*{CRediT authorship contribution statement}
\textbf{D. Ning:} Conceptualization, Investigation, Methodology, Writing – original draft, review \& editing.
\textbf{A. Dreizler:} Funding acquisition, Project administration, Writing – review \& editing.

\section{Introduction}
The combustion of nonvolatile metal particles or powders presents a promising strategy for achieving carbon-free conversion of clean energy on demand. Unlike the vapor-phase combustion observed in volatile droplets or the heterogeneous combustion of coal and biomass particles, where both mass and diameter typically decrease during combustion, nonvolatile metal particles exhibit a \textcolor{black}{distinctly} burning behavior. As the nonvolatile burning process progresses, the condense-phase reaction product accumulates within the particle, leading to an increase in both particle mass and diameter.
This unconventional combustion feature deviates significantly from the established theories developed for vapor-phase or heterogeneous combustion of droplets and volatile particles, such as the classical $d^2$-law \cite{glassman2014combustion}, which originated from the vaporization of liquid droplets. When applied to the scenario of nonvolatile particle combustion, these theories lack a rigorous physical foundation.
Recognizing the deficiency in understanding the combustion of nonvolatile particles, recent years have witnessed a surge in research efforts. Iron particles, in particular, have garnered substantial attention, with numerous studies aimed at unraveling the intricacies of their combustion behavior. \textcolor{black}{Other metals, like aluminium and silicon, are also expected to burn heterognerously at elevated pressures and reduced oxygen mole fractions, where the boiling points of the metals exceed their combustion temperatures.}

In experimental measurements, multiple concepts have been proposed to investigate the fundamentals of single iron particle combustion. Wright et al. \cite{wright2015combustion} measured the burn time of single iron particles by injecting single particles, ranging in diameter from 30 to $60\,$\textmu m, into preheated O$_2$/Ar mixtures at \SI{1000}{K} and atmospheric pressure. \textcolor{black}{It is oberved that the particle burn time (defined as the time to peak luminosity) reduces as the oxygen concentration increases.} Later, Ning et al. \cite{ning2021burn} further quantified the burn time of micron-sized iron particles, ignited by a focused laser beam and subsequently burning at room temperature. It is inferred that the combustion process from ignition to the peak temperature is limited by external oxygen diffusion, supported by the inverse proportionality between the burn time and the oxygen mole fraction in the ambient. Using the same particle generation and ignition apparatus, the evolution of the surface temperature and diameter during iron particle combustion were also quantified by Ning et al. \cite{ning2023size}. The measurement revealed that particle diameter increases almost linearly with elevating particle temperature, reaching a plateau during the cooling phase. This observation suggests that a relatively rapid oxidation rate is sustained throughout the particle burn time, with further oxidation occurring at a significantly slower rate after the particle peaks in temperature. In addition to laser ignition accomplished at low gas temperature, the combustion of iron particles in hot environments provided by the exhaust of lean-premixed gaseous flames \cite{li2022ignition,ning2023multi,toth2020combustion} or a drop-tube furnace \cite{panahi2023combustion} has been more extensively investigated. It is commonly observed that as the oxygen mole fraction increases, the burn time shortens, and simultaneously, the maximum particle temperature rises until a moderate oxygen level. These trends align with typical characteristics of particle combustion in the diffusion-dominated regime (but unnecessarily the diffusion-limited regime).

In numerical simulations, investigations into single iron particle combustion have been conducted from various perspectives, including particle-resolved modeling \cite{thijs2023resolved}, the point-particle approximation \cite{fujinawa2023combustion,xu2024phase}, and molecular-scale surface chemisorption \cite{thijs2023surface}. \textcolor{black}{Based on the assumption of the diffusion-limited combustion regime, Thijs et al. \cite{thijs2023resolved} performed quantitative modeling of burn time and temperature for laser-ignited iron particles, corresponding to experimental setups \cite{ning2021burn,ning2022temperature,ning2022critical}.}  The simulated burn time aligns well with measurements across a wide range of oxygen levels. However, calculated particle peak temperatures only match experimental results at relatively low oxygen mole fractions. At high oxygen levels, the numerical simulation tends to overestimate the experimental outcome. To account for this discrepancy, Fujinawa et al. \cite{fujinawa2023combustion} speculated that at relatively high oxygen levels, the rate-limiting mechanism for iron particle combustion may transition to the diffusion of ions in the liquid oxide layer.
Employing reactive molecular dynamic simulations, Thijs et al. \cite{thijs2023surface} demonstrated that the continuum model, constructed for the diffusion-limited regime, effectively describes the combustion behavior of iron particles with dimensions on the order of tens of microns up to the peak temperature not exceeding approximately \SI{2500}{K}. This finding indicates that conservation equations with the continuum assumption remain applicable for the theoretical analysis of nonvolatile particle combustion, provided that the particle size is significantly larger than the mean free path, and the mass accommodation coefficient of the oxygen is close to unity.

In theoretical frameworks, analytical solutions offer a comprehensive overview of general problems, complementing experiments and numerical simulations to enhance our understanding. Unfortunately, the quest for justified analytical models with predictive capability and quantitative accuracy for nonvolatile particle combustion remains elusive, impeded by the complexity of solving the nonlinear differential equations governing the problem.
Bidabadi and his collaborators  \cite{bidabadi2013analytical,maghsoudi2020analytical,bidabadi2013time} attempted theoretical analysis of single iron particle combustion using several mathematical approaches. However, these theoretical models were formulated on the basis of self-contradictory assumptions: 1) the oxidation rate, described by an Arrhenius-like expression, \textcolor{black}{was} assumed to be proportional to the particle surface area throughout the entire combustion process, a justification applicable only to the kinetic-limited regime; 2) the burn time \textcolor{black}{was} evaluated using an analytical solution derived for the lifetime of vapor-phase droplet combustion in the diffusion-limited regime \cite{glassman2014combustion}. Consequently, the results obtained from these models lack \textcolor{black}{a rigorous physical justification}. 
Accounting for the variation of particle diameter during oxidation, Hazenberg \cite{hazenberg2019eulerian} derived an analytical solution for the burn time of nonvolatile particles in the diffusion-limited combustion regime. While the expression explicitly describes the dependence of burn time on other parameters, %it can only be considered qualitative and incomplete because 
\textcolor{black}{for the mass diffusivity of oxygen, it} needs estimation from one-to-one numerical simulations or using measured particle temperatures from experiments \cite{ning2024experimental}. Moreover, the model becomes less rigorous with increasing oxygen levels due to the neglect of Stefan flow. Therefore, there is a need to further develop a comprehensive analytical theory for quantitatively predicting the combustion characteristics of nonvolatile particles in the diffusion-limited regime. This study aims to uncover the underlying objectives in this regard.

The paper is structured as follows. In \cref{sec: Mathematical formulation and solutions}, the interfacial heat and mass transfer rates are derived from the conservation equations of the gas phase \textcolor{black}{considering} relevant boundary conditions. Subsequently, the burn time of nonvolatile particle combustion in the diffusion-limited regime is formulated explicitly. 
The transient solution for the particle temperature is analytically solved using Taylor expansion and appropriate approximations. The coupling of the solutions for the burn time and time-averaged temperature of the particle yields quantitative results \textcolor{black}{for} defined boundary and initial conditions.
In \cref{sec: application}, the theory is applied to analyze the combustion of iron particles, considering various effects of particle and gas-phase properties. The theoretical results are then compared with experiments to validate the proposed theory. Finally, the main conclusions drawn from this study are summarized in \cref{sec: Conclusions}.

%develop a generalized, analytical model of nonvolatile particle combustion in the diffusion-limited regime, which is capable to describe the primary burning properties in a quantitatively accurate manner. 

%The current work aims to  

\section{Mathematical formulation and solutions}
\label{sec: Mathematical formulation and solutions}
\begin{figure}[h]
\centering\includegraphics[width=0.9\linewidth]{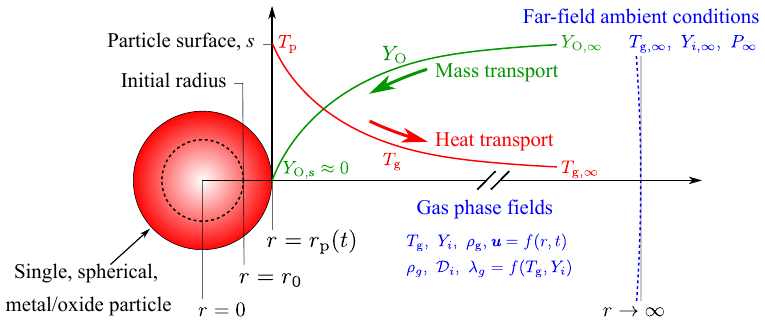}
\caption{Schematic illustration of the combustion process for a single, non-volatile metal particle in the diffusion-limited regime.}
\label{fig: burning particle illustration}
\end{figure}

The problem being addressed is illustrated in \cref{fig: burning particle illustration}. An isolated, spherical metal particle is suspended in a quiescent, isotropic, unconfined gaseous mixture. The ignition of the pure metal particle is simplified by initiating it at \textcolor{black}{its} ignition temperature. The subsequent nonvolatile combustion process is assumed to be limited by the external diffusion of the oxygen (i.e., the diffusion-limited regime), and the structure of the particle remains spherically symmetric, facilitating even chemisorption of the oxygen on the particle surface. 
During combustion, the heat generated by exothermic oxidation is transported outward, while the oxygen is \textcolor{black}{transported} inward. Meanwhile, the mass of the particle increases due to the incorporation of the oxygen, resulting in a growth of the particle size. As the particle temperature may exceed the melting points of the metal and \textcolor{black}{oxide}, the solid particle \textcolor{black}{may} transform into a liquid droplet consisting of both liquid metal and liquid oxide. Nevertheless, it will still be referred to as a particle in this context. For the sake of simplicity, the phase transitions of the particle are neglected, adopting constant specific heat. The temperature, $T_\mathrm{g}$, and composition, $Y_i$, of the gas vary between the particle surface and the far-field as $r \to \infty$, where $r$ is the radial coordinate. The combustion time and temperature history of the particle will be solved analytically.

\subsection{Interfacial heat and mass transfer rates}
 During the combustion of non-volatile fuel particles, gas-phase reactions are absent and the combustion product is formed on the particle surface. Therefore, employing the quasi-steady-state assumption, the gas phase conservation equations for mass, the mass fraction of the oxygen, and energy are written in a spherical coordinate system as
\begin{equation}
    \label{eqn: continuty}
    \frac{\partial}{\partial r} \left(r^2 \rho_\mathrm{g}  \bm{u}\right) = 0,
\end{equation}
\begin{equation}
    \label{eqn: species}
    \frac{\partial}{\partial r} \left(r^2 \rho_\mathrm{g}  \bm{u}Y_\mathrm{O}\right) - \frac{\partial}{\partial r}\left(r^2  \rho_\mathrm{g}\mathcal{D}_\mathrm{O} \frac{\partial Y_\mathrm{O}}{\partial r}\right) = 0,
\end{equation}
\begin{equation}
    \label{eqn: energy}
    \frac{\partial}{\partial r} \left(r^2 \rho_\mathrm{g} \bm{u}  h_\mathrm{g}\right) - \frac{\partial}{\partial r}\left(r^2 \lambda_\mathrm{g}  \frac{\partial T_\mathrm{g}}{\partial r}\right) - \frac{\partial}{\partial r} \left( \sum_{i=1}^{N}r^2 \rho_\mathrm{g} \mathcal{D}_i h_i  \frac{\partial Y_i}{\partial r}\right)= 0,
\end{equation}
where $\rho$, $\bm{u}$, $Y$, $\mathcal{D}$, $h$, and $\lambda$ represent the density, velocity, mass fraction, mass diffusion coefficient, specific enthalpy, and conductivity, respectively.  The subscripts, g, O, and $i$ denote the gas phase, oxygen, and individual species, respectively. $N$ is the number of species in the gas phase.
The boundary conditions at the particle surface are
\begin{subequations}
    \begin{empheq}[left={r=r_\mathrm{p}:\empheqlbrace}]{align}
        & \rho_\mathrm{g} \mathcal{D}_\mathrm{O} \frac{\partial Y_\mathrm{O}}{\partial r}\bigg|_\mathrm{s} -\rho_\mathrm{g} \bm{u}_\mathrm{s} Y_\mathrm{O,s} = \frac{\dot m_\mathrm{O}}{4\pi r_\mathrm{p}^2},
        \label{BC1}\\
        & Y_\mathrm{O} = Y_\mathrm{O,s},
        \label{BC2}\\
        & \lambda_\mathrm{g} \frac{\partial T_\mathrm{g}}{\partial r}\bigg|_\mathrm{s} =  h_\mathrm{T} (T_\mathrm{g,\infty} - T_\mathrm{p}) = \frac{\dot Q_\mathrm{conv}}{4\pi r_\mathrm{p}^2}, 
        \label{BC3}\\
        & T_\mathrm{g}  = T_\mathrm{p},
        \label{BC4}
    \end{empheq}
\end{subequations}
with $\dot m_\mathrm{O}$ being the mass consumption rate of the oxygen, $h_\mathrm{T}$ the heat transfer coefficient, and $\dot Q_\mathrm{conv}$ the convective heat transfer rate between the particle and the gas phase. The boundary conditions at the far-field read 
\begin{subequations}
    \begin{empheq}[left={r\to \infty:\empheqlbrace\,}]{align}
    & Y_\mathrm{O} = Y_\mathrm{O,\infty},\\
    & T_\mathrm{g}  = T_\mathrm{g, \infty}.
    \end{empheq}
\end{subequations}
To derive the interfacial heat and mass transfer rates, we will assume that the thermo-physical properties remain constant. This assumption is found to be satisfactory when the properties are evaluated at a virtual film layer with the reference conditions:
\begin{equation}
        T_\mathrm{f} := T_\mathrm{p} + A_\mathrm{f}(T_\mathrm{g,\infty}-T_\mathrm{p}),
    \end{equation}
    and
\begin{equation}
        Y_\mathrm{O,f} := Y_\mathrm{O,s} + A_\mathrm{f}(Y_\mathrm{O,\infty}-Y_\mathrm{O,s}),
        \label{averaging rule}
\end{equation}
with $A_\mathrm{f}$ being the weighting coefficient, commonly taken as 1/3 \cite{hubbard1975droplet}. 

We will first derive the interficial mass transfer rate of the oxygen. Since only the oxygen %(and the only) 
has a net mass flux,
integrating \eqref{eqn: continuty} from $r_\mathrm{p}$ to $\infty$ gives
\begin{equation}
\label{eqn: mass conservation}
     r^2 \rho_\mathrm{g} \bm{u} = \textcolor{black}{-} \frac{\dot m_\mathrm{O}}{4\pi},
\end{equation}
where the negative sign indicates that $\dot m_\mathrm{O}$ is inward.
Substituting \eqref{eqn: mass conservation} into \eqref{eqn: species} gives

\begin{equation}
    \frac{\partial}{\partial r} \left(-\frac{\dot m_\mathrm{O}}{4\pi}Y_\mathrm{O} - r^2  \rho_\mathrm{g} \mathcal{D}_\mathrm{O} \frac{\partial Y_\mathrm{O}}{\partial r} \right) =0.
    \label{eqn: oxygen consevation}
\end{equation}
Integrating \cref{eqn: oxygen consevation} from $r_\mathrm{p}$ to $r$, and applying the boundary condition \eqref{BC1} yields
\begin{equation}
    \label{eqn: oxygen conservation}
     4\pi r^2 \rho_\mathrm{g} \mathcal{D}_\mathrm{O} \frac{\text{d} Y_\mathrm{O}}{\text{d} r} = \dot m_\mathrm{O}(1- Y_\mathrm{O}).
\end{equation}
Integrating \cref{eqn: oxygen conservation} from $r$ to $\infty$ yields
\begin{equation}
    \begin{split}
        \ln \left(\frac{1-Y_\mathrm{O,\infty}}{1-Y_\mathrm{O}(r)} \right) = - \frac{1}{r}\frac{\dot m_\mathrm{O}}{4\pi\rho_\mathrm{g} \mathcal{D}_\mathrm{O}}.  \\
        %& = - 2\pi d_p \rho_g D_{O,m} \ln \left(1-Y_{O,\infty}\right).
        %&=   -\pi d_p (\rho_g D_{O,m})_f Sh B_{M},
    \end{split}
    \label{eq: oxygen distribution}
\end{equation}
Applying the boundary condition \eqref{BC2} to the above equation \eqref{eqn: oxygen conservation} gives the first expression for $\dot m_\mathrm{O}$:
 \begin{equation}
    \begin{split}
        \dot m_\mathrm{O} = -4\pi r_\mathrm{p}  \rho_\mathrm{g} \mathcal{D}_\mathrm{O}  \ln \left(1+B_\mathrm{M} \right), 
        %& = - 2\pi d_p \rho_g D_{O,m} \ln \left(1-Y_{O,\infty}\right).
        %&=   -\pi d_p (\rho_g D_{O,m})_f Sh B_{M},
    \end{split}
    \label{eq: 1st mass transport rate}
\end{equation}
 with $B_\mathrm{M}$ being the Spalding mass transfer number of the oxygen, defined analogously to that of vaporized fuel \citep{spalding1960standard} as
 
\begin{equation}
    B_\mathrm{M} := \frac{Y_\mathrm{O,\infty}-Y_\mathrm{O,s}}{Y_\mathrm{O,s}-1}.
    \label{oxident spalding number}
\end{equation}
We introduce the Sherwood number, Sh,  defined as
\begin{equation}
    \text{Sh} := 2\frac{\ln \left( 1+B_\mathrm{M}\right)}{B_\mathrm{M}},
    \label{sherwood number}
\end{equation}
which inherently takes the Stefan flow effect into account. Except for the atmosphere with (nearly) pure oxygen \citep[p. 506]{glassman2014combustion}, in the diffusion-limited combustion regime:
\begin{equation}
    Y_\mathrm{O,s} \approx 0 \ \Rightarrow \ B_\mathrm{M} \approx -Y_\mathrm{O,\infty}. 
    \label{eqn: simple BM}
\end{equation}
Substituting \eqref{sherwood number} and \eqref{eqn: simple BM} into \eqref{eq: 1st mass transport rate} 
yields
\begin{equation}
    \begin{split}
        \dot m_\mathrm{O} & = \text{Sh}   \pi d_\mathrm{p} \rho_\mathrm{g} \mathcal{D}_\mathrm{O}  Y_\mathrm{O,\infty}
         = -2 \pi d_\mathrm{p} \rho_\mathrm{g} \mathcal{D}_\mathrm{O}  \ln(1-Y_\mathrm{O,\infty}).
    \end{split}
    \label{eq: Sh mass transport rate}
\end{equation}
The explicit expression of the interficial heat transfer rate, $\dot Q_\mathrm{conv}$, can be derived by following a similar procedure as for $\dot m_\mathrm{O}$.
We consider the facts that the net mass flux of each inert  species is zero:
\begin{equation}
    \label{eqn: enthalpy inert}
     r^2 \rho_\mathrm{g} \bm{u}  Y_i  -  r^2 \rho_\mathrm{g} \mathcal{D}_\mathrm{i} \frac{\partial Y_i}{\partial r} = 0,\ i\in [1,N]\ \&\ i\neq i_\mathrm{O}
\end{equation}
\begin{comment}
We consider the facts that the net mass flux of all inert species is zero:
\begin{equation}
    \label{eqn: enthalpy inert}
     r^2 \rho_\mathrm{g} \bm{u} \sum_{i=1, i\neq \mathrm{O}}^{N} Y_i  -  \sum_{i=1, i\neq \mathrm{O}}^{N}r^2 \rho_\mathrm{g} \mathcal{D}_\mathrm{i} \frac{\partial Y_i}{\partial r} = 0,
\end{equation}
\end{comment}
and the specific enthalpy of the gaseous mixture encompasses those of individual species:
 \begin{equation}
    \label{eqn: enthalpy of gas}
     h_\mathrm{g} = \sum_{i=1}^{N} h_i  Y_i = h_\mathrm{O} Y_\mathrm{O} +  \sum_{i=1, i\neq i_\mathrm{O}}^{N}  h_i Y_i.
 \end{equation}
Substituting \cref{eqn: mass conservation,eqn: oxygen conservation,eqn: enthalpy inert,eqn: enthalpy of gas} into \cref{eqn: energy} yields
\begin{equation}
\label{eqn: simplified energy equation}
    \frac{\partial}{\partial r} \left( - \frac{\dot m_\mathrm{O}}{4\pi} h_\mathrm{O} - \lambda_\mathrm{g} r^2 \frac{\partial T_\mathrm{g}}{\partial r} \right) = 0.
\end{equation}
Integrating \cref{eqn: simplified energy equation} from $r_\mathrm{p}$ to $r$ and applying the boundary condition \eqref{BC3} gives
\begin{equation}
    \label{eqn: simplified energy 2}
     4 \pi r^2 \lambda_\mathrm{g} \frac{\text{d} T_\mathrm{g}}{\text{d} r} = \dot m_\mathrm{O}c_{p,\mathrm{O}} \left( T_\mathrm{p} -  T_\mathrm{g} +  \frac{\dot Q_\mathrm{conv}}{\dot m_\mathrm{O} c_{p,\mathrm{O}}} \right),
\end{equation}
where the relation $h = c_pT$ has been used. Integrating \cref{eqn: simplified energy 2} from $r$ to $\infty$ gives $T_\mathrm{g}$ as a function of the radial distance:
\begin{equation}
    \ln \left(\frac{T_\mathrm{p}-T_\mathrm{g,\infty}\, + \frac{\dot Q_\mathrm{conv}}{\dot m_\mathrm{O} c_{p,\mathrm{O}}}}{T_\mathrm{p}-T_\mathrm{g}(r) + \frac{\dot Q_\mathrm{conv}}{\dot m_\mathrm{O} c_{p,\mathrm{O}}}} \right) = -\frac{1}{r}\frac{\dot m_\mathrm{O}c_{p,\mathrm{O}}}{4 \pi\lambda_\mathrm{g}}
    \label{eqqn:gas temperature distribution}
\end{equation}
At $r=r_\mathrm{p}$, \cref{eqqn:gas temperature distribution} provides the second expression for $\dot m_\mathrm{O}$:
\begin{equation}
    \label{eqn: 2nd mass consumption rate}
    \dot m_\mathrm{O} = - 4\pi r_\mathrm{p} \frac{\lambda_\mathrm{g}}{c_{p,\mathrm{O}}}\ln(1+B_\mathrm{T}),
\end{equation}
where $B_\mathrm{T}$ is the Spalding heat transfer number in our case, defined as
\begin{equation}
\label{eqn: heat transfer number}
    B_\mathrm{T} := \frac{\dot m_\mathrm{O}c_{p,\mathrm{O}}}{\dot Q_\mathrm{conv}}(T_\mathrm{p} - T_\mathrm{g,\infty}).
\end{equation}
Substituting \cref{eqn: 2nd mass consumption rate} into \cref{eqn: heat transfer number} yields

\begin{equation}
    \dot Q_\mathrm{conv} = 4 \pi r_\mathrm{p}  \lambda_\mathrm{g}  \frac{\ln(1+B_\mathrm{T})}{B_\mathrm{T}}  (T_\mathrm{g,\infty} - T_\mathrm{p}).
    \label{heat transfer rate}
\end{equation}
We introduce the Nusselt number, Nu, defined as
\begin{equation}
    \text{Nu}:=2\frac{\ln(1+B_\mathrm{T})}{B_\mathrm{T}}.
    \label{eqn: Nu number}
\end{equation}
Substituting \cref{eqn: Nu number} into \cref{heat transfer rate} gives
\begin{equation}
    \dot Q_\mathrm{conv} = \text{Nu} \pi d_\mathrm{p} \lambda_\mathrm{g} (T_\mathrm{g,\infty} - T_\mathrm{p})
    \label{heat transfer rate with Sh}
\end{equation}
Finally, the two expressions for $\dot m_\mathrm{O}$, i.e., \cref{eq: 1st mass transport rate,eqn: 2nd mass consumption rate}, relate $B_\text{T}$ and $B_\text{M}$:
\begin{equation}
   1+ B_\mathrm{T}=(1+B_\mathrm{M})^{\frac{1}{\mathrm{Le_O}}\frac{c_{p,\mathrm{O}}}{c_{p,\mathrm{g}}}},
   \label{eq: relation between BT and BM}
\end{equation}
where $\mathrm{Le_O} = \lambda_\mathrm{g}/(\rho_\text{g} c_{p,\text{g}} \mathcal{D}_\mathrm{O}$) is the Lewis number of the oxygen at the film layer. In the follows, the mass diffusivity of the oxygen and the thermal conductivity and specific heat of the gaseous mixture  evaluated at the film layer will be denoted as  $\left(\rho_g\mathcal{D}_\mathrm{O} \right)_\mathrm{f}$, $\lambda_\mathrm{f}$, and $c_{p,\mathrm{f}}$, respectively.
\if
As aforementioned, in \cref{eq: Sh mass transport rate,heat transfer rate with Sh}, $\rho_gD_{O,m}$ and $\lambda_g$ should be evaluated at the film layer:

\begin{equation}
    \left\{
    \begin{aligned}
        &\left(\rho_gD_{O,m} \right)_f = \rho_g (T_f, Y_{i,f}) \cdot D_{O,m}(T_f, Y_{i,f})\\
        &\lambda_{g,f} = \lambda_g(T_f,Y_{i,f})
    \end{aligned} 
    \right.
\end{equation}
\fi

%the interfacial mass and heat transfer rate of burning non-volatile particles, which largely resembles the classical theory of droplets evaporation.

\subsection{Analytical solution for particle burn time}
During non-volatile combustion, the particle mass increases by consuming the oxygen,  which is described as
\begin{equation}
    \frac{\mathrm{d}m_\mathrm{p}}{\mathrm{d}t} = \frac{\pi}{6}\frac{\mathrm{d}\left(\rho_\mathrm{p} d_\mathrm{p}^3\right)}{\mathrm{d}t} = \dot m_\mathrm{O},
\label{particle mass governing equation}
\end{equation}
with $m_\mathrm{p}$ and $\rho_\mathrm{p}$ being the time-dependent mass and density of the particle, respectively. As derived in \ref{App:A} (see Supplementary Material), $\rho_\mathrm{p}$ is a function of the density of the metal, $\rho_\mathrm{m}$, and that of the metal oxide, $\rho_\mathrm{mo}$:
\begin{equation}
\label{eqn: time-depandant particle diameter}
   \rho_\mathrm{p} =\rho_1 + \rho_2\left(\frac{d_0}{d_\mathrm{p}}\right)^3,
\end{equation}
with
\begin{equation}
    \rho_1 := \frac{\rho_\mathrm{mo}}{s+(1-s)\varrho},
\end{equation}
and
\begin{equation}
    \rho_2 := \frac{(\rho_\mathrm{m}-\rho_\mathrm{mo})s}{s+(1-s)\varrho},
\end{equation}
where $s$ is the stoichiometric oxide-to-oxygen mass ratio, and $\varrho$ is the oxide-to-metal density ratio.
Substituting \cref{eqn: time-depandant particle diameter,eq: Sh mass transport rate} into \cref{particle mass governing equation} yields
\begin{equation}
    \frac{\mathrm{d}\left(\rho_1d_\mathrm{p}^3 + \rho_2d_0^3 \right)}{\mathrm{d}t} = -12 d_\mathrm{p} \left(\rho_\mathrm{g} \mathcal{D}_\mathrm{O}\right)_\mathrm{f}  \ln \left( 1-Y_\mathrm{O,\infty} \right).
    \label{governing eqn 2}
\end{equation}
Because $\rho_2d_0^3$ is constant, \cref{governing eqn 2} simplifies as
\begin{equation}
    \rho_1 d_\mathrm{p} \frac{\mathrm{d} \left( d_\mathrm{p} \right)}{\mathrm{d}t} = -4 \left(\rho_\mathrm{g} \mathcal{D}_\mathrm{O}\right)_\mathrm{f} \ln \left( 1-Y_\mathrm{O,\infty} \right).
    \label{governing eqn 3}
\end{equation}
Integrating \cref{governing eqn 3} from $t_0$ ($t_0=0$: $d_\mathrm{p} = d_0$) to $t$ ($t_0\leqslant t \leqslant t_\mathrm{b}$) gives the particle diameter evolution:
\begin{equation}
    d_\mathrm{p}^2 = d_0^2 -\frac{8 \left( \rho_\mathrm{g} \mathcal{D}_\mathrm{O} \right)_\mathrm{f}\ln \left( 1-Y_\mathrm{O,\infty} \right)}{\rho_1}t.
\label{size evolution}
\end{equation}
where $d_0$ is the initial particle diameter.
Since the particle size grows during combustion, we define an \textit{expansion coefficient}:
\begin{equation}
    K := -\frac{8 \left( \rho_\mathrm{g} \mathcal{D}_\mathrm{O} \right)_\mathrm{f} \ln \left( 1-Y_\mathrm{O,\infty} \right)}{\rho_1}.
\end{equation}
Consequently, \cref{size evolution} becomes
\begin{equation}
    d_\mathrm{p}^2 = d_0^2 + Kt.
    \label{size evolution with K}
\end{equation}
\Cref{size evolution with K} resembles the size evolution of a shrinking droplet during vaporization (i.e., $d^2 = d_0^2 - \beta t$ with $\beta$ the \textit{evaporation coefficient} \cite{glassman2014combustion}). 
Substituting the final condition of combustion  (i.e., $t=t_\mathrm{b}$: $d_\mathrm{p} = d_1$ and $d_1$ is the final particle diameter) into \cref{size evolution} gives the analytical solution for the burn time:
\begin{equation}
    t_\mathrm{b} = \frac{\rho_1\left( d_0^2-d_1^2 \right)}{8 \left( \rho_\mathrm{g} \mathcal{D}_\mathrm{O} \right)_\mathrm{f}\ln \left( 1-Y_\mathrm{O,\infty} \right)}
    = \frac{\rho_1\left( 1-\epsilon^2\right)d_0^2}{8 \left(\rho_\mathrm{g} \mathcal{D}_\mathrm{O}\right)_\mathrm{f}\ln \left( 1-Y_\mathrm{O,\infty} \right)},
    \label{diffusion-limited burn time}
\end{equation}
with $\epsilon:=d_1/d_0$ defining the diameter expansion ratio.
\Cref{diffusion-limited burn time} indicates a $d_0^2$-dependence of the burn time in the diffusion-limited regime.
Besides, using the time-dependent density  and diameter of the particle, given in \cref{eqn: time-depandant particle diameter} and \cref{size evolution with K} respectively, the transient increase of the particle mass is determined as
\begin{equation}
    m_\mathrm{p} = \frac{\pi}{6}\left[\rho_1 (d_0^2 + Kt)^{3/2} + \rho_2 d_0^3  \right].
    \label{eqn: particle mass history}
\end{equation}
Substituting \cref{{eqn: particle mass history}} into \cref{mass metal} and normalize the result by the initial particle mass gives the fraction of remaining unburned metal in the particle during combustion:
\begin{equation}
    \frac{m_\mathrm{m}}{m_0} = (1-s)\left[\frac{\rho_1}{\rho_\mathrm{m}} \left(1 + \frac{K}{d_0^2}t\right)^{3/2} + \frac{\rho_2}{\rho_\mathrm{m}}\right]  + s.
    \label{eqn: unburnt mass}
\end{equation}
%where $\rho_\mathrm{F}$ is the density of the metal.

\subsection{Analytical solution for transient particle temperature}
\textcolor{black}{For some metals (e.g., Fe),  a majority of the combustion process is in the liquid phase, where the specific heats of the metal and the oxide are close and independent of temperature \cite{chase1998nist}.  Therefore, the specific heat of the particle does not change significantly during combustion.}
Assuming a constant specific heat of the particle, the  governing equation of the particle temperature is expressed as
\begin{equation}
    c_{p,\mathrm{p}}\frac{\mathrm{d}(m_\mathrm{p}T_\mathrm{p})}{\mathrm{d}t} = 
    c_{p,\mathrm{p}}m_\mathrm{p}\frac{\mathrm{d}T_\mathrm{p}}{\mathrm{d}t} + c_{p,\mathrm{p}}T_\mathrm{p}\frac{\mathrm{d}m_\mathrm{p}}{\mathrm{d}t}
    =
    \dot Q_\mathrm{conv} + \dot Q_\mathrm{chem} + \dot Q_\mathrm{rad},
\label{Tp governing equation 1}
\end{equation}
with $c_{p,\mathrm{p}}$ being the specific heat of the particle, 
$m_\mathrm{p} = \rho_\mathrm{p}\pi d_\mathrm{p}^3/6$ the particle mass, $\dot Q_\mathrm{conv}$ the convective heat transfer rate, $\dot Q_\mathrm{chem}$ the chemical heat release rate, and $\dot Q_\mathrm{rad}$ the radiative heat loss rate. Since only nonvolatile combustion is considered in this work, the heat loss via evaporation is neglected. Although evaporation may become important when the particle temperature approaches the boiling point of the metal or the oxide, it is beyond the scope of strictly-defined nonvolatile combustion and makes the analytical solution \textcolor{black}{seemingly} impossible to complete. Consequently, the results derived here may be limited to scenarios where the maximum particle temperature is considerably below the boiling points.
\begin{comment}
After applying the chain rule, \cref{Tp governing equation 1} writes
\begin{equation}
    c_{p,\mathrm{p}}m_\mathrm{p}\frac{\mathrm{d}T_\mathrm{p}}{\mathrm{d}t} + c_{p,\mathrm{p}}T_\mathrm{p}\frac{\mathrm{d}m_\mathrm{p}}{\mathrm{d}t}  = \dot Q_\mathrm{conv} + \dot Q_\mathrm{chem} + \dot Q_\mathrm{rad} + \dot Q_\mathrm{vap}.
\label{Tp governing equation 2}
\end{equation}

The particle mass reads
\begin{equation}
    m_\mathrm{p} = \rho_\mathrm{p}\frac{\pi d_\mathrm{p}^3 }{6},
\label{particle mass}
\end{equation}
where $\rho_\mathrm{p}$ and $d_\mathrm{p}$ are the time-dependent particle density and diameter, respectively.
\end{comment}
Combining \cref{eq: Sh mass transport rate,particle mass governing equation} gives the rate of particle mass change in the diffusion-limited combustion regime: 
\begin{equation}
    \frac{\mathrm{d}m_\mathrm{p}}{\mathrm{d}t}  = \dot m_\mathrm{O} =  \mathrm{Sh} (\rho_\mathrm{g} \mathcal{D}_\mathrm{O})_\mathrm{f} \pi d_\mathrm{p} Y_\mathrm{O,\infty},
\label{particle mass change rate}
\end{equation}
Rewriting \cref{heat transfer rate with Sh}, the convective heat transfer rate reads
\begin{equation}
\label{eq: heat convection rate}
   \dot{Q}_\mathrm{conv} = \mathrm{Nu}\lambda_\mathrm{f}\pi d_\mathrm{p} (T_\mathrm{g,\infty}-T_\mathrm{p}).
\end{equation}
%with Nu the Nusselt number, $\lambda_\mathrm{f}$ the conductivity of gas at the film layer, $d_\mathrm{p}$ particle diameter, $T_\mathrm{g}$ and $T_\mathrm{p}$ the temperatures of gas and particle, respectively.
The chemical heat release rate is written as
\begin{equation}
\label{eq: chemical heat release}
   \dot{Q}_\mathrm{chem} = \dot{m}_\mathrm{O_2} \phi \Delta h,
\end{equation}
where $\phi$ is the stiochiometric fuel-to-oxygen mass ratio, and $\Delta h$ is the combustion enthalpy per gram fuel. The radiative heat transfer rate is described by the Stefan-Boltzmann law:
\begin{equation}
\label{eq: radiative heat loss}
    \dot Q_\mathrm{rad} = - \pi d_\mathrm{p}^2 \varepsilon\sigma(T_\mathrm{p}^4 - T_0^4),
\end{equation}
%\textcolor{black}{\subsection{Without radiation and vaporization}}
%\label{subsec: Without radiation and vaporization}
where $\varepsilon$ is the emissivity of the particle surface and $\sigma$ is the Stefan-Boltzmann constant.
Substituting \cref{particle mass change rate,eq: heat convection rate,eq: chemical heat release,eq: radiative heat loss} into \cref{Tp governing equation 1} and rearranging it yields:
\begin{equation}
    \begin{split}
    %    c_{p,s} \frac{\rho_p d_p^2}{6} \frac{dT_p}{dt}  = Nu\lambda_f (T_g - T _p) + Sh    
    %    (\rho D)_f Y_{O, \infty}(\phi \Delta h - c_{p, s}T_p)
        c_{p,\mathrm{p}} \frac{\rho_\mathrm{p} d_\mathrm{p}^2}{6} \frac{\mathrm{d}T_\mathrm{p}}{\mathrm{d}t} 
        &= \mathrm{Nu}\lambda_\mathrm{f}T_\mathrm{g,\infty} + \mathrm{Sh} (\rho_\mathrm{g} \mathcal{D}_\mathrm{O})_\mathrm{f} Y_\mathrm{O, \infty} \phi \Delta h \\ 
        & - \left( \mathrm{Nu}\lambda_\mathrm{f} + \mathrm{Sh} (\rho_\mathrm{g} \mathcal{D}_\mathrm{O})_\mathrm{f} Y_\mathrm{O, \infty} c_{p,\mathrm{p}} \right) T_\mathrm{p} \\
        & - d_\mathrm{p} \varepsilon\sigma(T_\mathrm{p}^4 - T_0^4).
    \end{split}
\label{eqn: Tp governing equation 2}
\end{equation}
With  $\rho_\mathrm{p} = \rho_1 + \rho_2(d_0/d_\mathrm{p})^3$ (\cref{eqn: time-depandant particle diameter}) and $d_\mathrm{p}^2 = d_0^2 + Kt$ (\cref{size evolution with K}), the following equation is obtained:
\begin{equation}
    \rho_\mathrm{p} d_\mathrm{p}^2 = \rho_1 d_\mathrm{p}^2  + \rho_2 \frac{d_0^3}{d_\mathrm{p}} = \rho_1 (d_0^2 + Kt) + \rho_2 \frac{d_0^3}{\sqrt{d_0^2 + Kt}}.
\label{eqn: density diameter square}
\end{equation}
In order to solve \cref{eqn: Tp governing equation 2} analytically, \cref{eqn: density diameter square} needs to be approximated. To this end, we consider the Taylor expansion at $t=0$ for the second term on the right-hand side of \cref{eqn: density diameter square}, which expresses as
\begin{equation}
    \frac{d_0^3}{\sqrt{d_0^2 + Kt}} = \frac{d_0^2}{\sqrt{1 + (K/d_0^2)t}} =  \sum_{n=0}^{\infty} \left( \frac{K}{d_0^2} t \right)^n \frac{(2n-1)!!}{n!(-2)^n} d_0^2.
    \label{eqn: nonlinear term}
\end{equation}
Recalling \cref{size evolution with K,density ratio,density ratio with particle diameter}, one can find the following inequality valid for many metals with examples given in \cref{tab: exmaples}, where thermal expansion is neglected:
\begin{equation}
\label{eq: tylor expasion}
    \left|\frac{K}{d_0^2} t \right| = \frac{d_\mathrm{p}^2}{d_0^2} - 1 \le \frac{d_1^2}{d_0^2} - 1 = \left(\frac{\rho_\mathrm{m}\mathcal{M}_\mathrm{mo}}{\rho_\mathrm{mo}\mathcal{M}_\mathrm{m}} \frac{1}{x} \right)^{2/3} - 1 < 1,\, \forall t \in [0,t_\mathrm{b}],
\end{equation}
in which $\mathcal{M}_\mathrm{m}$ and $\mathcal{M}_\mathrm{mo}$ are the molar mass of the metal and the metal oxide, respectively, $x$ is the numbers of metal atom in the chemical formula of the metal oxide, M$_x$O$_y$. Therefore, higher order terms in \cref{eq: tylor expasion} become smaller.
By neglecting $\mathcal{O}(t^2)$ %$\textit{O}(t^2)$
\footnote{It is possible to find a solution when the second order is included but the improvement is negligibly small and the expression is cumbersome. Therefore, we do not present it in the paper.}, 
\cref{eqn: nonlinear term} is approximated as
\begin{equation}
    \frac{d_0^3}{\sqrt{d_0^2 + Kt}} \approx  d_0^2  - \frac{K}{2}t,\, \forall t \in [0,t_\mathrm{b}].
    \label{eq: linear approximation of the 2nd term}
\end{equation}
%----------------------------------------------------------------------------------------
\iffalse
Therefore, we can neglect $\mathcal{O}(t^3)$ and \cref{eqn: nonlinear term} is approximated as
\begin{equation}
    \frac{d_0^3}{\sqrt{d_0^2 + Kt}} \approx  d_0^2  - \frac{K}{2}t + \frac{3K^2}{8}t^2, \forall t \in [0,t_b].
\end{equation}
\fi
%----------------------------------------------------------------------------------------
Then, \cref{eqn: density diameter square} is also approximated as
\begin{equation}
    \begin{split}
        \rho_\mathrm{p} d_\mathrm{p}^2 & \approx \rho_1 (d_0^2 + Kt) + \rho_2 ( d_0^2  - \frac{K}{2}t)\\
        & = (\rho_1 + \rho_2) d_0^2 + \left(\rho_1 - \frac{\rho_2}{2} \right)Kt
    \end{split}
\label{eqn: density diameter square approximation}
\end{equation}
The maximum truncation errors in \cref{eq: linear approximation of the 2nd term,eqn: density diameter square approximation} introduced by the neglect of higher order terms in the Taylor series, \cref{eqn: nonlinear term}, reach at $t_\mathrm{b}$, which do not exceed 5\% as provided in \cref{tab: exmaples} for several sample metals.
Substituting \cref{eqn: density diameter square approximation} into \cref{eqn: Tp governing equation 2} and reaggranging it yields
\begin{equation}
    \begin{split}
    %    c_{p,s} \frac{\rho_p d_p^2}{6} \frac{dT_p}{dt}  = Nu\lambda_f (T_g - T _p) + Sh    
    %    (\rho D)_f Y_{O, \infty}(\phi \Delta h - c_{p, s}T_p)
        \left(1 + \frac{2\rho_1 -\rho_2}{2(\rho_1 +\rho_2)d_0^2}K t \right)\frac{\mathrm{d}T_\mathrm{p}}{dt} &\approx \textcolor{black}{6}\frac{\text{Nu}\lambda_\mathrm{f}T_\mathrm{g,\infty} + \text{Sh} (\rho_\mathrm{g} \mathcal{D}_\mathrm{O})_\mathrm{f} Y_\mathrm{O, \infty} \phi \Delta h}{c_{p,\mathrm{p}}(\rho_1 +\rho_2)d_0^2} \\ 
         &  - \textcolor{black}{6}\frac{ \mathrm{Nu}\lambda_\mathrm{f} + \text{Sh} (\rho_\mathrm{g} \mathcal{D}_\mathrm{O})_\mathrm{f} Y_\mathrm{O, \infty} c_{p, \mathrm{p}} }{c_{p,\mathrm{p}}(\rho_1 +\rho_2)d_0^2} T_\mathrm{p} \\
        & - \textcolor{black}{6} \frac{ d_\mathrm{p} \varepsilon\sigma(T_\mathrm{p}^4 - T_0^4)}{c_{p,\mathrm{p}}(\rho_1 +\rho_2)d_0^2}
    \end{split}
\label{eqn: Tp governing equation approximation}
\end{equation}
By defining constant parameters as follows:
\begin{equation}
\label{eqn: constant B}
    \mathfrak{B} := \frac{2\rho_1 -\rho_2}{2(\rho_1 +\rho_2)d_0^2}K,
\end{equation}

\begin{equation}
\label{eqn: constant D}
    \mathfrak{D} := \textcolor{black}{6}\frac{\mathrm{Nu}\lambda_\mathrm{f}T_\mathrm{g,\infty} + \mathrm{Sh} (\rho_\mathrm{g} \mathcal{D}_\mathrm{O})_\mathrm{f} Y_\mathrm{O, \infty} \phi \Delta h}{c_{p,\mathrm{p}}(\rho_1 +\rho_2)d_0^2},
\end{equation}

\begin{equation}
\label{eqn: constant E}
    \mathfrak{E} := \textcolor{black}{6}\frac{ \mathrm{Nu}\lambda_\mathrm{f} + \mathrm{Sh} (\rho_\mathrm{g} \mathcal{D}_\mathrm{O})_\mathrm{f} Y_\mathrm{O, \infty} c_{p, \mathcal{p}} }{c_{p,\mathcal{p}}(\rho_1 +\rho_2)d_0^2},
\end{equation}
and a non-linear function:
\begin{equation}
    \mathbf{F} (T_\mathrm{p}) := \textcolor{black}{6} \frac{ d_\mathrm{p} \varepsilon\sigma(T_\mathrm{p}^4 - T_0^4)}{c_{p,\mathrm{p}}(\rho_1 +\rho_2)d_0^2}
     \approx \textcolor{black}{6} \frac{ \varepsilon\sigma(T_\mathrm{p}^4 - T_0^4)}{c_{p,\mathrm{p}}(\rho_1 +\rho_2)d_0}  
     \label{eq: non-linear F}
\end{equation}
in which the approximation is obtained by replacing $d_\mathrm{p}$ with $d_0$. Alternatively, $d_\mathrm{p}$ can also be replaced by $d_1$. The difference in the solved maximum particle temperatures that will be discussed in \cref{sec: application} is only around 1\% between these two approximations, which can be regarded as negligible. The exact solution is bounded by those solved using these two approximations.
Substituting \cref{eqn: constant B,eqn: constant D,eqn: constant E,eq: non-linear F} into
\cref{eqn: Tp governing equation approximation} simplifies it as
\begin{equation}
    (1+\mathfrak{B}t)\frac{\mathrm{d}T_\mathrm{p}}{\mathrm{d}t} \approx \mathfrak{D}-\mathfrak{E}T_\mathrm{p} - \mathbf{F} (T_\mathrm{p}).
    \label{Tp governing equation simplified}
\end{equation}
Rearranging  \cref{Tp governing equation simplified} and integrating both sides yields
\begin{equation}
    \int \frac{\mathrm{d}T_\mathrm{p}}{\mathfrak{D}-\mathfrak{E}T_\mathrm{p} - \mathbf{F} (T_\mathrm{p})} \approx \int \frac{\mathrm{d}t}{1+\mathfrak{B}t }
    \label{intergal of Tp governing equation}
\end{equation}
To derive an explicit solution for $T_\mathrm{p}(t)$ from \cref{intergal of Tp governing equation}, we first neglect radiation, i.e., $\mathbf{F} (T_\mathrm{p}) = 0$. Then the general solution of \cref{intergal of Tp governing equation} expresses as
\begin{equation}
\label{eqn: particle temperatue solution}
    T_\mathrm{p}(t) \approx k_1(1 + \mathfrak{B}t)^{\mathfrak{-E/B}} + \frac{\mathfrak{D}}{\mathfrak{E}},
\end{equation}
with $k_1$ the constant of integration, which can be determined as
\begin{equation}
\label{eqn: k1}
    k_1 = T_\mathrm{p,ign} - \frac{\mathfrak{D}}{\mathfrak{E}},
\end{equation}
using  the initial condition:
\begin{equation}
\label{eqn: inital condition}
    t = 0:\ T_\mathrm{p}(0) = T_\mathrm{p,ign}.
\end{equation}
Finally, the explicit solution of \cref{eqn: Tp governing equation approximation} reads
\begin{equation}
\label{eqn: particle temperatue solution with radiation}
    T_\mathrm{p}(t) \approx \left(T_\mathrm{p,ign} - \frac{\mathfrak{D}}{\mathfrak{E}}\right)(1 + \mathfrak{B}t)^{\mathfrak{-E/B}} + \frac{\mathfrak{D}}{\mathfrak{E}}.
\end{equation}

%However, it is also possible that the particle peak temperature is limited by the burnout of the metal before researching the steady state. The reasonable solution for the peak temperature shpuld be   

%with 
%\begin{equation}
%    \frac{\mathfrak{D}}{\mathfrak{E}} = \frac{\mathrm{Nu}\lambda_\mathrm{f}T_\mathrm{g} + %\mathrm{Sh} (\rho_\mathrm{g} \mathcal{D}_\mathrm{O})_\mathrm{f} Y_\mathrm{O, \infty} \phi %\Delta h}{\mathrm{Nu}\lambda_\mathrm{f} + \mathrm{Sh} (\rho_\mathrm{g} %\mathcal{D}_\mathrm{O})_\mathrm{f} Y_\mathrm{O, \infty}},
%\end{equation}

%in order to solve \cref{intergal of Tp governing equation}, the time-dependent particle diameter in $\mathbf{F} (z)$ can be approximated by $d_0$ or $d_1$. The difference in the solved maximum particle temperatures that will be discussed in \cref{sec: application} is only around 1\% between these two approximations, which can be regarded negligible.

When radiation is considered, 
an explicit solution cannot be found but $T_\mathrm{p}$ can be expressed implicitly as
\begin{equation}
\label{eq: Tp with radiation}
    t \approx \frac{1}{\mathfrak{B}} \left[ \exp \left( \int_{T_\mathrm{p,ign}}^{T_\mathrm{p}(t)}\frac{\mathfrak{B}\text{d}z}{\mathfrak{D}-\mathfrak{E}z - \mathbf{F} (z)}  \right) - 1 \right].
\end{equation}
The integral in \cref{eq: Tp with radiation} can be solved analytically using partial fraction decomposition, yielding
\begin{equation}
    \int_{T_\mathrm{p,ign}}^{T_\mathrm{p}(t)}\frac{\mathfrak{B}\text{d}z}{\mathfrak{D}-\mathfrak{E}z - \mathbf{F} (z)} = \mathfrak{B}\sum_{\omega_i} \frac{\ln\left(T_\mathrm{p,ign} - \omega_i \right) - \ln\left(T_\mathrm{p} - \omega_i \right)}{4\mathfrak{F}\omega_i + \mathfrak{E}},
\end{equation}
where $\omega_i$ is the complex roots of the quartic equation $\mathfrak{D}-\mathfrak{E}\omega - \mathbf{F} (\omega) = 0$ and $\mathfrak{F}$  is a constant defined as
\begin{equation}
    \mathfrak{F}:=\textcolor{black}{6} \frac{\varepsilon\sigma}{c_{p,\mathrm{p}}(\rho_1 +\rho_2)d_0}
\end{equation}

\begin{table}[h]
    \centering
    \caption{Values of \(d_{1}^2/d_{0}^2 -1\) for selected metals and oxides, along with the corresponding truncation errors (TE) induced by the linear approximation in \cref{eq: linear approximation of the 2nd term,eqn: density diameter square approximation}, respectively.}
    \label{tab: exmaples}
    \begin{tabular}{ccccc}
    \hline
    Metal & Oxide  & $d_1^2/d_0^2 -1$  & TE in \cref{eq: linear approximation of the 2nd term} & TE in \cref{eqn: density diameter square approximation}\\
    \hline
    Al & Al$_2$O$_3$ & 0.19  & -1.3\% & 1.3\%\\
    Si & SiO$_{2}$   & 0.52  & -8.8\% & 1.2\%\\
    Fe & FeO         & 0.46  & -7.0\% & -3.4\%\\
    \hline
    \end{tabular}
\end{table}

\subsection{Coupling of solutions and computational procedure}

The analytical solutions for particle burn time and temperature evolution have been derived separately. However they are conjugated via the mass diffusivity of the oxygen at the film layer, $\left(\rho_\mathrm{g} \mathcal{D}_\mathrm{O}\right)_\mathrm{f}$. The mass diffusivity is unknown \textit{a priori} since it is evaluated at the film layer temperature, $T_\mathrm{f}$, whose determination needs the particle temperature as shown in \cref{averaging rule}. Since the particle temperature is transient, so is the film layer temperature. 

\begin{figure}[h!]
\centering\includegraphics[width=0.5\linewidth]{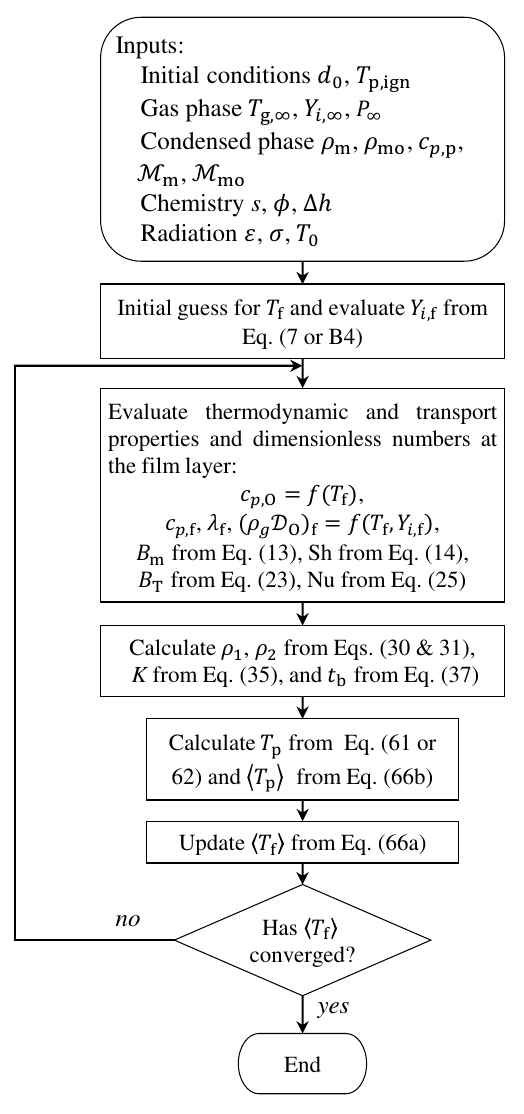}
\caption{Calculation procedure for obtaining the quantitative solution of the burn time and temperature history . }
\label{fig: calculation procedure}
\end{figure}

Considering the fact that $\left(\rho_\mathrm{g} \mathcal{D}_\mathrm{O}\right)_\mathrm{f} \propto T_\mathrm{f}^{0.7}$  for diatomic gases, which is close to be a linear correlation, the following approximate relation is obtained:
\begin{equation}
   \left(\rho_g \mathcal{D}_\mathrm{O}\right)_\mathrm{f} \appropto \left(1-A_\mathrm{f}\right)T_\mathrm{p} + A_\mathrm{f} T_\mathrm{g,\infty}
\end{equation}
This approximation implies that, on the one hand, the variation of $\left(\rho_\mathrm{g} \mathcal{D}_\mathrm{O}\right)_\mathrm{f}$ during particle combustion is not significant owing to a constant far-field gas temperature. On the other hand, due to the linear proportionality, an averaged $\left(\rho_\mathrm{g} \mathcal{D}_\mathrm{O}\right)_\mathrm{f}$ could be evaluated at a time-averaged film layer temperature:
\begin{subequations}
\begin{equation}
    \left<T_\mathrm{f}\right> = (1-A_\mathrm{f})\left<T_\mathrm{p}\right> + A_\mathrm{f} T_\mathrm{g,\infty}
\end{equation}
with $\left<T_p\right>$ the time-averaged particle temperature:
\begin{equation}
    \left<T_\mathrm{p}\right> = \frac{1}{t_\mathrm{b}}\int_0^{t_\mathrm{b}}T_\mathrm{p}(t)\mathrm{d}t 
\end{equation}
\label{eq: time-averaged film-layer temperature}
\end{subequations}
To obtain quantitative results, 
%the analytical solutions for the particle burn time and temperature need to be re-coupled. This is because the governing equations for the mass (\cref{particle mass governing equation}) and temperature (\cref{Tp governing equation 1}) of the particle are linked via the mass transfer rate of the oxygen (\cref{eq: Sh mass transport rate}) that is affected by the particle temperature via the average rule, \cref{averaging rule}.
the time-average temperature and gas composition at the film layer are determined iteratively using the procedure illustrated in \cref{fig: calculation procedure}.
The thermal physical and transport properties of the gaseous mixture and species at the film layer are evaluated using Cantera \cite{cantera}.

%To determine $\lambda_f$ and $(\rho D)_f$ , the temperature and gas composition in the film layer needs to be estimated using commonly adopted averaging rules:

\section{Application to iron particle combustion}
\label{sec: application}

In this section, the theory developed above is applied to analyze the burning properties of micron-sized iron particles, which serves as a typical example of non-volatile, heterogeneous combustion at low to moderate oxygen concentrations. This choice is justified by the much higher boiling points of iron and iron oxides ($>\,$$3000\,$K at \SI{1}{atm}) compared to the peak temperature of the burning particles in air at room temperature ($\sim\,$$2500\,$K \cite{ning2022temperature}). Additionally, systematic experiments providing quantitative data have been conducted, offering an opportunity to validate the theory. 

\subsection{Iron-oriented assumptions and parameters}

The basic assumptions, which were adopted to reasonably simplify the practical problem, leading to the \textcolor{black}{successful derivation} of the theory, such as constant thermophysical properties and homogeneous distributions of temperature inside the particle, remain unchanged. 
The main assumptions related to iron particle combustion are three-fold.
First, the ignition temperature is assumed to be \SI{1100}{K}, as suggested by the calculations using a state-of-the-art model \cite{mi2022quantitative}.
Second, the oxidation reaction is assumed to be $\text{Fe} + \text{O}_2 \to \text{FeO}$, and combustion enthalpy per gram iron is determined by the NASA thermodynamic database. 
Third, the reaction stops upon completely conversion to FeO, and further oxidation is neglected.
The physical parameters used for performing the quantitative analysis of iron particle combustion are listed in \cref{tab: constant parameters for iron combustion}.

\begin{table}[h]
\centering
\caption{Physical parameters for iron particle combustion.}
\begin{tabular}{lll}
\hline
 Parameter & Value  & Unit \\
 \hline
 $c_{p,\mathrm{p}}$ & 0.94 & $\mathrm{J}\cdot \mathrm{K}^{-1}\cdot \mathrm{g}^{-1}$\\
 $\rho_\mathrm{Fe}$ & 7.874 & $\mathrm{g}\cdot \mathrm{cm}^{-3}$\\
 $\rho_\mathrm{FeO}$ & 5.740 & $\mathrm{g}\cdot \mathrm{cm}^{-3}$\\
 $\phi$ & 3.5 & -  \\
 $\Delta h$ & 4875 & $\mathrm{J}\cdot \mathrm{g}^{-1}$\\
\hline
\end{tabular}
\label{tab: constant parameters for iron combustion}
\end{table}

\subsection{Theoretical results and discussion}
In this section, sample results on the combustion characteristics of iron particles under different boundary and initial conditions will be presented. The mechanisms underling the theoretical results are explained. %, often from a mathematical perspective. Finally, the theoretical calculations are compared to experiments, and the results are validated through this comparison.
%The burning characteristics of micro-metric single iron particles have experimentally studied in two common ways. In the first way, the particle is rapidly heated by a laser beam to or above the ignition temperature usually at room temperature for simplicity. The others approach is to release individual particles into a hot oxidizing atmosphere.  In the next, we will theoretically examine single iron particle combustion corresponding to these two experimental ways.
%If the particle combustion is dominated by O$_2$ diffusion to the particle surface, the burning process continues until the burnout. Whereas, the kinetic-controlled 
%One distinct feature of the diffusion-limited combustion of metal particles is that they can burn at room temperature, whereas this is impossible in the kinetic-limited regime. 

\subsubsection{Effects of particle size and surface radiation}

\begin{figure}[h!]
\centering\includegraphics[width=0.7\linewidth]{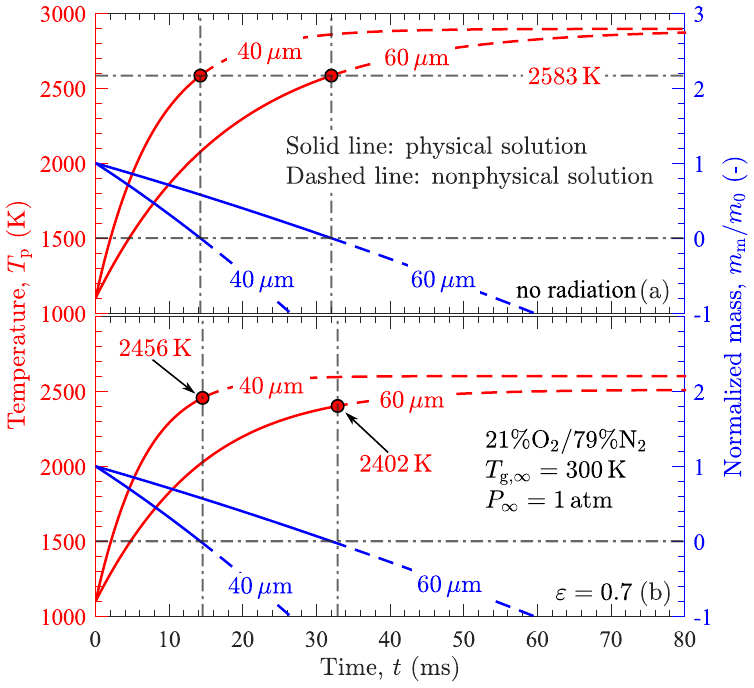}
\caption{Theoretical temperature history and fraction of unburned iron during the combustion of a $40\,$\textmu m and a $60\,$\textmu m iron particles burning in the $21\%\text{O}_2/79\%\text{N}_2$ mixture at $T_\mathrm{g,\infty}=300\,$K and $P_{\infty}=1\,$ atm, (a) without radiation and (b) with radiation ($\varepsilon = 0.7$).}
\label{fig: sample results}
\end{figure}

\Cref{fig: sample results} illustrates the quantitative solutions for the temperature and the fraction of unburnt iron of a  \SI{40}{}{\,}\textmu m and a $60\,$\textmu m iron particles during combustion in the 21\%\ce{O2}$/$79\%\ce{N2}  at \SI{300}{K}, both with and without considering radiation. 
When the radiative heat loss is neglected, the two particles reach the same peak temperature of \SI{2578}{K} at approximately \SI{14}{ms} and \SI{32}{ms}, respectively. By the same time, iron has burnt out, as indicated by the disappearance of unburnt iron mass in \cref{fig: sample results}(a). 
The independence of the particle peak temperature on the diameter arises because $d_0$ cancels out from the analytical solution for the particle peak temperature, as derived in \cref{eqn: particle peak temperatue solution}. This solution is obtained by substituting the particle burn time solution, \cref{diffusion-limited burn time}, into the radiation-excluded expression of particle temperature history, \cref{eqn: particle temperatue solution}.
\begin{equation}
\label{eqn: particle peak temperatue solution}
    \begin{split}
    T_\mathrm{p, max} & \approx \left(T_\mathrm{p,ign} - \frac{\mathfrak{D}}{\mathfrak{E}}\right)
     \left( 1 + \frac{(2\rho_1-\rho_2)(1-\epsilon^2)}{2(\rho_1+\rho_2)} \right)^{-\mathfrak{E}/\mathfrak{B} }
      + \frac{\mathfrak{D}}{\mathfrak{E}}.
    \end{split}
\end{equation}
As time continues to elapse, the particle temperature still increases until reaching a plateau, corresponding to the steady-state solution for the particle temperature. This plateau represents the time-independent term, $\frac{\mathfrak{D}}{\mathfrak{E}}$, in \cref{eqn: particle temperatue solution}. The steady-state solution is achieved when the rates of the conductive heat loss and the chemical heat release are balanced.
However, this solution is nonphysical because it yields a negative value for the unburnt mass of iron. 
When radiation is considered, the dependence of the particle peak temperature on the particle diameter emerges, indicating that the larger particle has a lower peak temperature. The maximum temperatures of the small and large particles are \SI{2456}{K} and \SI{2402}{K}, about \SI{60}{K} and \SI{90}{K} lower than their non-radiative counterparts, respectively. Additionally, \cref{fig: sample results} shows that the effect of radiative heat loss on particle burn times is negligible. The overall \textcolor{black}{temporal evolution} of the particle temperature, including both the physical and nonphysical solutions, clearly depicts that the maximum particle temperature is constrained by the burnout of fuel rather than heat losses.

\begin{figure}[h!]
\centering\includegraphics[width=0.7\linewidth]{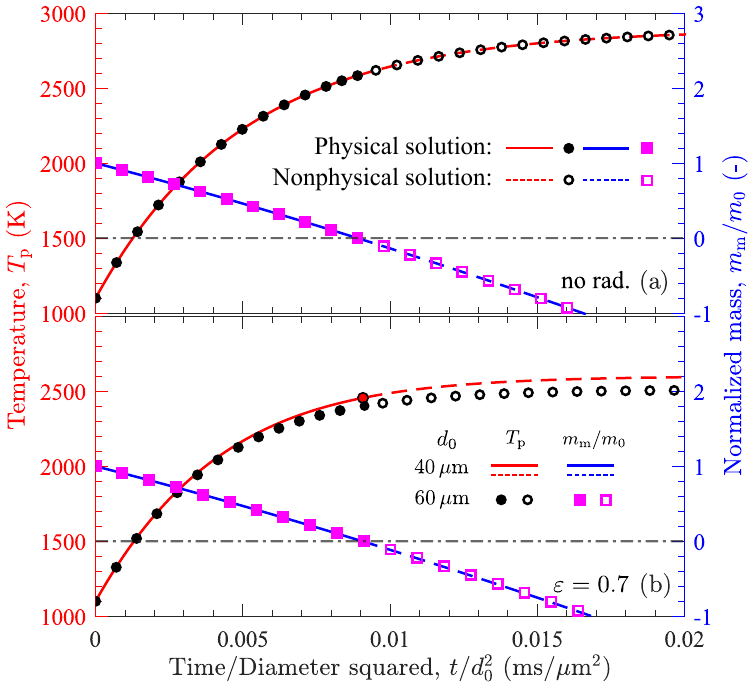}
\caption{Scaled temperature history and fraction of unburned iron during the combustion of a $40\,$\textmu m and a $60\,$\textmu m iron particles burning at the room condition: $21\%\text{O}_2\slash79\%\text{N}_2$, $T_\mathrm{g,\infty}=300\,$K, and $P_\mathrm{\infty}=1\,$atm, (a) without radiation and (b) with radiation ($\varepsilon = 0.7$).}
\label{fig: sample results scaled time}
\end{figure}

%
\begin{comment}
The effect of radiation is more profound on the nonphysical steady-state solution of the particle temperature, making the maximum particle temperature is more close the the steady-state solution, \SI{2601}{K} and \SI{2508}{K}, for the small and the big particles, respectively. This will help us to explain the physical reason for the size dependence of the maximum particle temperature in the follows.
Since we do not have an explicit solution for the particle peak temperature when radiation is considered, an intuitive explanation for the size dependence of the particle peak temperature could be made approximately based on the the steady-state solution. In the steady-state, both the convective heat loss rate $\dot Q_\mathrm{conv}$ and the chemical heat release rate $\dot Q_\mathrm{chem}$ scale with the particle diameter as shown by \cref{eq: heat convection rate,eq: chemical heat release}, respectively. Whereas, the radiative heat loss scales with the particle diameter squared. Consequently, the steady-state particle temperature (i.e., $T_\mathrm{p,ss}$) is related to the radiative heat loss as
\begin{equation}
    T_\mathrm{p, ss} \propto \frac{\dot Q_\mathrm{rad}}{d_\mathrm{p}} = - \pi d_\mathrm{p} \varepsilon\sigma(T_\mathrm{p, ss}^4 - T_0^4),\ \forall T_\mathrm{p,ss}>T_\mathrm{0}.
\end{equation}
The above expression suggests that the bigger the particle, the lower the steady-state particle temperature. 
\end{comment}
%
The time history of the particle temperature and the fraction of unburnt iron shown in \cref{fig: sample results} are re-plotted versus the time coordinated normalized by the initial particle diameter squared (i.e., $t/d_0^2$) in \cref{fig: sample results scaled time}. %It is seen that the 
Curves for different particle diameters overlap when radiation is neglected. As the initial particle diameter, $d_0$, drops out in $\mathfrak{D/E}$ and $\mathfrak{E/B}$, the effect of the initial particle diameter on the particle temperature, described by \cref{eqn: particle temperatue solution}, \textcolor{black}{appears} only in the term:
 \begin{equation}
    \mathfrak{B}t = \frac{2\rho_1 -\rho_2}{2(\rho_1 +\rho_2)}K \cdot \frac{t}{d_0^2}.
    \label{term Bt}
\end{equation}
As the first factor of the right hand side of \cref{term Bt} is independent of $d_0$, the particle temperature history scales with $1/d_0^2$ when radiation is neglected. This scaling behavior is also valid for the fraction of unburnt iron, as described by \cref{eqn: unburnt mass}. Even when accounting for radiation, \textcolor{black}{the curves for 40 and 60$\,$\textmu m coincide very well.} For particle peak temperatures below approximately $\sim$\SI{2500}{K}, radiation appears to have a weak impact on controlling the temperature evolution of burning iron particles, particularly for particles with relatively small diameters (e.g., $<100\,$\textmu m). \textcolor{black}{This scaling is consistent with the expanding $d^2$-law derived in \cref{size evolution with K}.} 

%
\begin{comment} 
In other words, when the temperature trajectories of burning particles with different initial diameters, given by \cref{eqn: particle temperatue solution}, 
In other words, when the temperature trajectories of burning particles with different initial diameters, given by \cref{eqn: particle temperatue solution}, are plotted versus the time coordinated normalized by the initial particle diameter squared (i.e., $t/d_0^2$), they overlap as shown in fig. XXX. Besides, the last term of \cref{eqn: particle temperatue solution} is the steady-state solution for the maximum particle temperature without considering radiation, which is due to the thermal balance between the conductive heat loss and heat release. 
\end{comment}

\subsubsection{Effects of gas-phase pressure and temperature}
\begin{figure}
  \centering\includegraphics[width=0.7\linewidth]{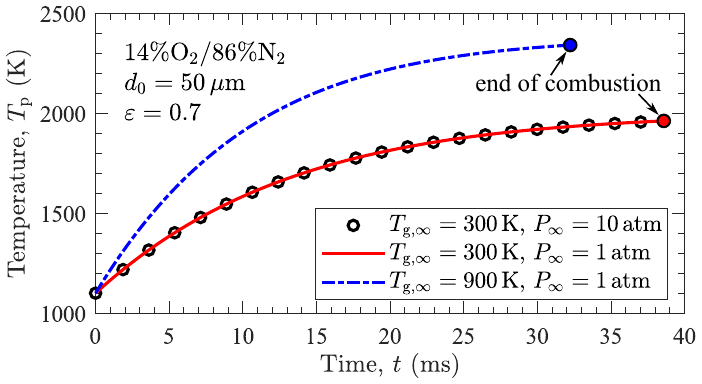}
  \caption{Time histories of temperature for a $50\,$\textmu m particle when burning at different oxygen mole fractions and ambient pressures. The oxygen mole fraction of 14\% is used to limit the maximum particle temperature below \SI{2500}{K} at the ambient temperature of \SI{900}{K}. }
  \label{fig: gas T and P}
\end{figure}

\Cref{fig: gas T and P} depicts the burn time and temperature history of a $50\,$\textmu m iron particle under varying gas-phase pressure, $P_\mathrm{\infty}$, and temperature, $T_\mathrm{\infty}$.
Increasing the ambient pressure from \SI{1}{atm} to \SI{10}{atm} exhibits no discernible impact on the diffusion-limited combustion process of the particle.   The independence of particle burn time from ambient pressure is attributed to the negligible influence of pressure on the mass diffusivity of the oxygen at the film layer, denoted as $\left(\rho_\mathrm{g}\mathcal{D}_\mathrm{O}\right)_\mathrm{f}$, a critical factor controlling the duration of particle combustion in the diffusion-limited regime. \textcolor{black}{The gas density, $\rho_\mathrm{g}$, scales linearly with pressure and the binary diffusion coefficient of oxygen, $\mathcal{D}_\mathrm{O}$, is inversely proportional to pressure according to Chapman-Enskog theory \cite{hirschfelder1954molecular}.}
\begin{comment}
    This phenomenon can be explained through the relationship between gas density, $\rho_\mathrm{g}$, and pressure, where the former is directly proportional to the latter. Meanwhile, the mass diffusion coefficient of the oxygen in a binary mixture (e.g., $\ce{O2}/\ce{N2}$), denoted as $\mathcal{D}_\mathrm{O}$, follows an inverse proportionality to pressure, as outlined in Chapman-Enskog’s theory \cite{hirschfelder1954molecular}:
\begin{equation}
    \mathcal{D}_{12} = \frac{3}{16}\frac{\sqrt{2\pi k^3_\mathrm{B}T^3/\mathcal{M}_{12}}}{P\pi\sigma^2_{12}\Omega}, 
\end{equation}
where  $k_\mathrm{B}$ denotes the Boltzmann constant, $\mathcal{M}_{12} = \mathcal{M}_1\mathcal{M}_2/(\mathcal{M}_1+\mathcal{M}_2)$ represents the reduced molecular mass for the species pair (subscripts 1 and 2 denote the two kinds of molecules in the gaseous mixture),
$\sigma_{12}$ is the reduced collision diameter, and $\Omega$ is the collision
integral. 
\end{comment}
Additionally, the thermal conductivity of ideal gases remains independent of pressure, except at extremely low pressures where the mean free path approaches or surpasses the dimensions of fuel particles (Knudsen number Kn $\nleqslant$ 1). Consequently, the governing equation for particle temperature (\cref{eqn: Tp governing equation approximation}) and its solutions (\cref{eqn: particle temperatue solution with radiation,eq: Tp with radiation}), depicting the time history of the particle temperature, are unaffected by ambient gas pressure.

\begin{figure}[h]
  \centering\includegraphics[width=0.9\linewidth]{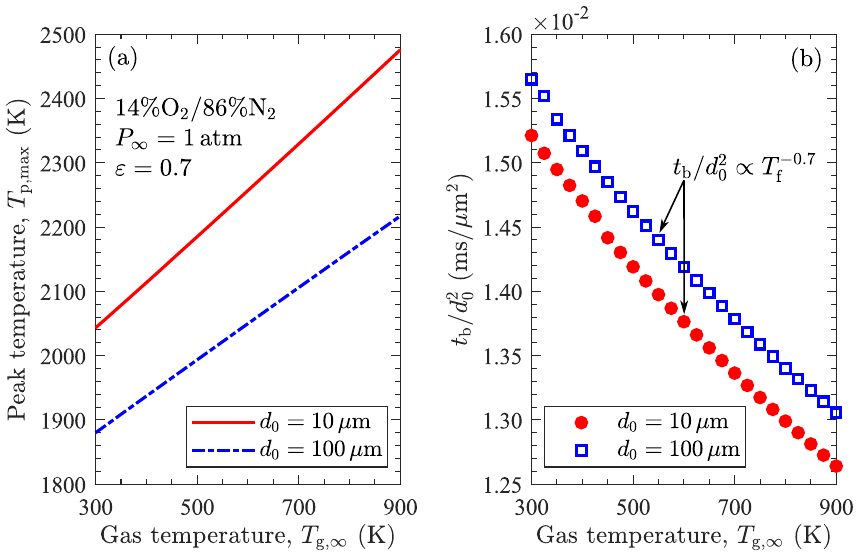}
  \caption{Particle peak temperature (a) and burn time scale with diameter squared (b) as a function of the far-field gas temperature.}
  \label{fig: Tmax and tmas vs Tg}
\end{figure}

In contrast to pressure, the burn time of the particle shortens with increasing the gas-phase temperature, ending at a higher maximum temperature. \Cref{fig: Tmax and tmas vs Tg}(a) summarizes the peak particle temperature, $T_\mathrm{p,max}$, and burn time normalized by the initial particle diameter squared, $t_\mathrm{b}/d_0^2$, as a function of the gas-phase temperature for 10 and $100\,$\textmu m iron particles burning in the 14\%\ce{O2}/86\%\ce{N2} mixture. The calculation is conducted for particle peak temperatures below \SI{2500}{K}, ensuring negligible evaporation. As shown in \Cref{fig: Tmax and tmas vs Tg}(a), the peak temperature shows an almost linear correlation with the gas-phase temperature. Given that the role of radiation is significantly weaker in determining the maximum particle temperature compared to convection, the linear dependence of the particle peak temperature on the far-field gas temperature can be derived by rearranging \cref{eqn: particle peak temperatue solution}, yielding 

%For \ce{O2}/\ce{N2} mixtures,  $\mathrm{Le_O} \approx 1$ and $c_{p,\mathrm{O}}=c_{p,\mathrm{g}}$. Therefore, $B_\mathrm{T}\approx B_\mathrm{M}$ according to \cref{eq: relation between BT and BM} and thus $\mathrm{Sh} \approx \mathrm{Nu}$
\begin{subequations}
\label{eq: linear dependence of Tpmax on Tg}
\begin{equation}
    \begin{split}
    T_\mathrm{p, max} &\approx (1-\mathfrak{G})\frac{\mathrm{Nu Le_O}c_{p,\mathrm{f}}}{\mathrm{Nu Le_O}c_{p,\mathrm{f}} + \mathrm{Sh} Y_\mathrm{O,\infty}c_{p,\mathrm{p}}}T_\mathrm{g,\infty}\\
    &+ (1-\mathfrak{G})\frac{Y_\mathrm{O,\infty}\phi\Delta h}{\mathrm{Nu Le_O}c_{p,\mathrm{f}} + \mathrm{Sh}Y_\mathrm{O,\infty}c_{p,\mathrm{p}}}\\ & + T_\mathrm{p,ign}\mathfrak{G} ,
    \end{split}
\end{equation}
with the $T_\mathrm{g,\infty}$-independent constant $\mathfrak{G}$ defined as
\begin{equation}
    \mathfrak{G} := \left( 1 + \frac{(2\rho_1-\rho_2)(1-\epsilon^2)}{2(\rho_1+\rho_2)} \right)^{\frac{3}{4}\frac{\mathrm{NuLe_O}c_{p,\mathrm{f}} + \mathrm{Sh}Y_\mathrm{O,\infty}c_{p,\mathrm{p}}}{\left(2-\rho_2/\rho_1 \right)\ln\left(1-Y_\mathrm{O,\infty}\right)}},
\end{equation}
\end{subequations}
where Le$_\mathrm{O}$ and $c_{p,\mathrm{g}}$ are the Lewis number of the oxygen and the specific heat of the gaseous mixture evaluated at the film layer, respectively. In \cref{eq: linear dependence of Tpmax on Tg}, all the other parameters are independent on the far-field gas temperature, $T_\mathrm{g,\infty}$, and thus the linearity between $T_\mathrm{p,max}$ and $T_\mathrm{g,\infty}$ is determined by the first term on the right hand side of the equation. 

In \cref{fig: Tmax and tmas vs Tg}(b), it is observed that as the far-field gas temperature increases, the particle burn time decreases in proportion to $T_\mathrm{f}^{-0.7}$. This behavior is attributed to the fact that the mass diffusivity of the oxygen at the film layer is directly proportional to $T_\mathrm{f}^{0.7}$, which increases with the rise in ambient gas temperature. Additionally, \textcolor{black}{the larger particle exhibits a larger value of $t_\mathrm{b}/d_0^2$ at the same gas temperature}. This can be attributed to increased heat loss through surface radiation. The heightened surface radiation leads to a reduction in the time-averaged particle temperature and, consequently, the time-averaged film-layer temperature, $\left<T_\mathrm{f}\right>$. This reduction in $\left<T_\mathrm{f}\right>$ contributes to a decrease in the mass diffusivity of the oxygen, resulting in larger values of $t_\mathrm{b}/d_0^2$.

\subsubsection{Effect of initial particle temperature}
\label{sec: Effect of ignition temperature}
Laser heating is a commonly employed experimental technique for initiating combustion in metal particles. Despite the advantages of being able to flexibly adjust the gas composition and temperature in laser-ignition experiments, there is a tendency for the particle to rapidly reach a high-temperature state in the diffusion regime without undergoing the typical ignition process. Since oxidation during laser heating is considered negligible, the resulting laser-driven high-temperature state can be treated as an initial particle temperature for theoretical analysis. This section examines the impact of the initial particle temperature on the combustion characteristics of iron particles.

\begin{figure}[h]
  \centering\includegraphics[width=0.7\linewidth]{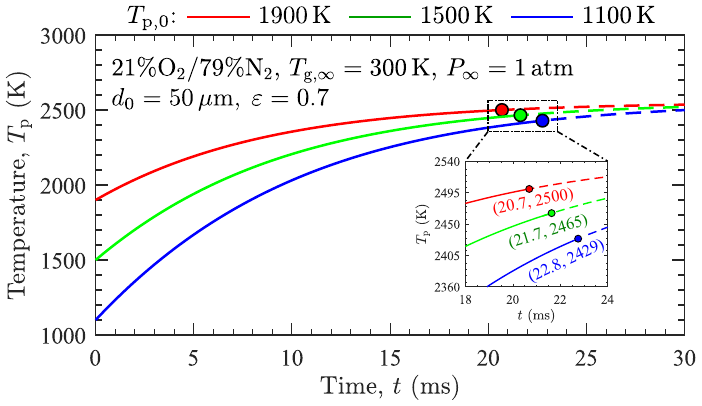}
  \caption{Time histories of a $50\,$\textmu m iron particle with three different initial temperature, burning in the mixture of $21\%\ce{O2}/79\%\ce{N2}$ at $T_\mathrm{g,\infty}=300\,$K, and $P_\mathrm{\infty}=1\,$atm.
  %The initial particle temperature triggered by laser has weak effect of the burn time and peak temperature of the particle, which depends on the initial particle size and oxygen concentration in the gas phase. For biggr particles, the effect is smaller because radiative cooling reduces the peak temperature. At lower oxygen concentration, the effect is weaker because the at burn out, particle is more close to the thermal equilibrium state where the initial sate has been relaxed.
  }
  \label{fig: Tp history with different T0}
\end{figure}

\Cref{fig: Tp history with different T0} shows the calculated time histories for different initial temperature of a $50\,$\textmu m iron particle burning in a mixture of 21\%\ce{O2}/79\%\ce{N2} at \SI{300}{K} and atmospherical pressure. Despite a temperature difference of \SI{400}{K} between neighboring cases at the onset of combustion, the maximum particle temperatures at burnout are very close, with a difference of about \SI{35}{K}. This weak effect of the initial temperature on the maximum particle temperature is attributed to the combustion progress closely approaching a steady state, where the initial condition has been largely relaxed.
Moreover, the significant elevations of the initial particle temperature only induce very minor reductions (i.e., $<10\%$) in the particle burn time. In the diffusion-limited regime, the insensitivity of the burn time to the initial particle temperature is due to the controlling parameter of particle burn time being the mass diffusivity of the oxygen $\rho_\mathrm{g}\mathcal{D}_\mathrm{O}$, evaluated using a time-averaged film layer temperature $\left<T_\mathrm{f}\right>$. According to \cref{eq: time-averaged film-layer temperature}, the role of the initial particle temperature in determining $\rho_\mathrm{g}\mathcal{D}_\mathrm{O}$ is significantly weakened by the two-fold averaging operations. 
For example, the calculated $\left<T_\mathrm{f}\right>$ in the cases of $T_\mathrm{p,0}=1100\,$K and \SI{1900}{K} shown in \cref{fig: Tp history with different T0} are \SI{1423}{K} and \SI{1638}{K}, respectively. This means that an 73\% variation in $T_\mathrm{p,0}$ causes only a 15\% change in $\left<T_\mathrm{f}\right>$. Since $\rho_\mathrm{g}\mathcal{D}_\mathrm{O} \propto \left<T_\mathrm{f}\right>^{0.7}$, the difference in the mass diffusivity of the oxygen further reduces to 9.7\%, which finally determines the discrepancies between the particle burn times.
The weak coupling between the burn time and temperature of the particle is a key feature of heterogeneous combustion in the diffusion-limited regime.

\subsection{Comparison with experiments}
To validate the predictive capability of the theory, the theoretical burn times and temperatures of micron-sized iron particles are compared with a series of measurements reported in Refs. \cite{ning2021burn,ning2023size,ning2022temperature}. These measurements were conducted for isolated iron particles burning in \ce{O2}/\ce{N2} mixtures at room temperature. As the particle temperature right after laser ignition could not be directly measured for most cases in the experiments \cite{ning2021burn,ning2022temperature}, an initial partial temperature of \SI{1500}{K} is assumed, identical to that used in Ref. \cite{thijs2023resolved}.

\begin{figure}[h]
\centering\includegraphics[width=0.9\linewidth]{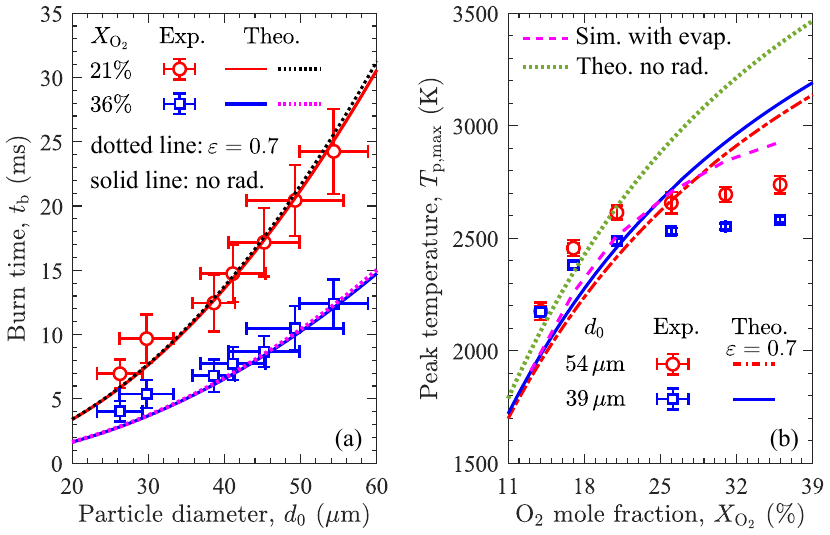}
\caption{(a) Comparison between experimental and theoretical burn times of iron particles with various initial diameters. (b) Experimental, theoretical, and numerical particle peak temperature as a function oxygen molar fractions. The experimental results are obtained for $39\pm2.8\,$\textmu m and $54\pm4.5\,$\textmu m particle size distributions. 
The theoretical results are calculated for the arithmetic mean diameters of $39\,$\textmu m and $54\,$\textmu m, and the simulation results are for a $50\,$\textmu m particle, adopted from Ref. \cite{thijs2023resolved}. }
\label{fig: tb and Tmax}
\end{figure}

\Cref{fig: tb and Tmax}(a) reveals a comparison between the theoretical and experimental burn times of laser-ignited iron particles within the size range of 25--$55\,$\textmu m. 
The calculated burn times, based on the current theoretical model, exhibit almost perfect agreement with the measurements for initial particle diameters larger than $38\,$\textmu m. Whereas for smaller particles, the theory slightly underestimates the burn time. This high level of agreement validates the quantitative accuracy of the current theoretical model in predicting particle burn times, not only at normal but also at elevated oxygen levels. 
Additionally, \cref{fig: tb and Tmax}(b) compares the particle peak temperatures predicted by the theory with those approximately measured using a spectrometer \cite{ning2022temperature}. In the experiment, the arithmetic mean diameters of the adopted particle size distributions are $39\,$\textmu m and $54\,$\textmu m, respectively. For the theoretical calculation, the mean diameters are used. The results show that at oxygen mole fractions below approximately 26\%, the theory reasonably captures the maximum particle temperature, although the measurements slightly surpass the calculations that include radiation. This discrepancy is likely attributed to the overestimation of the experimental approach, as the maximum particle temperatures are deduced from the measured gray-body emission spectra.  As explained in Ref. \cite{ning2021burn}, these spectra are prone to be dominated by the emissions of relatively hot particles appearing within the time period of data acquisition due to the strong dependence of thermal radiation on temperature. 
The maximum particle temperatures calculated without considering radiation are marginally higher than their radiation-included counterparts, especially at relatively low oxygen mole fractions. However, this discrepancy becomes more significant with increasing oxygen levels because the particle peak temperature rises, making radiative heat loss more prominent. 
Furthermore, the current theoretical results closely resemble particle-resolved direct numerical simulations for a $50\,$\textmu m particle \cite{thijs2023resolved}, which also incorporates the surface evaporation of liquid iron and oxide. When the mole fraction of oxygen exceeds 26\%, both the numerical simulation and the present theory remarkably overpredict the particle peak temperature. \textcolor{black}{As stated previously, evaporation is neglected in the theory, leading to the overpredication of the particle temperature at relatively high oxygen levels. In addition, this overestimation can also be attributed to the transition of the rate-limiting mechanism from external diffusion to either surface chemisorption \cite{thijs2023surface} or internal diffusion of ions across the liquid oxide layer \cite{fujinawa2023combustion}, which are neither considered in the theory nor the numerical simulation of Thijs et. al \cite{thijs2023resolved}.}  
Given the excellent agreement between the measured burn time and the calculation based on the assumption of the external-diffusion-limited combustion regime, it is plausible that the other two potential rate-limiting mechanisms may only come into play for a brief period near the end of the liquid-phase combustion of iron particles, specifically in proximity to the particle peak temperature. For the majority of the iron particle combustion process, it is likely that external diffusion of the oxygen predominantly dictates the dynamics, making the inner structure of the particle less influential for predicting burn time.

\begin{figure}[h!]
\centering\includegraphics[width=0.7\linewidth]{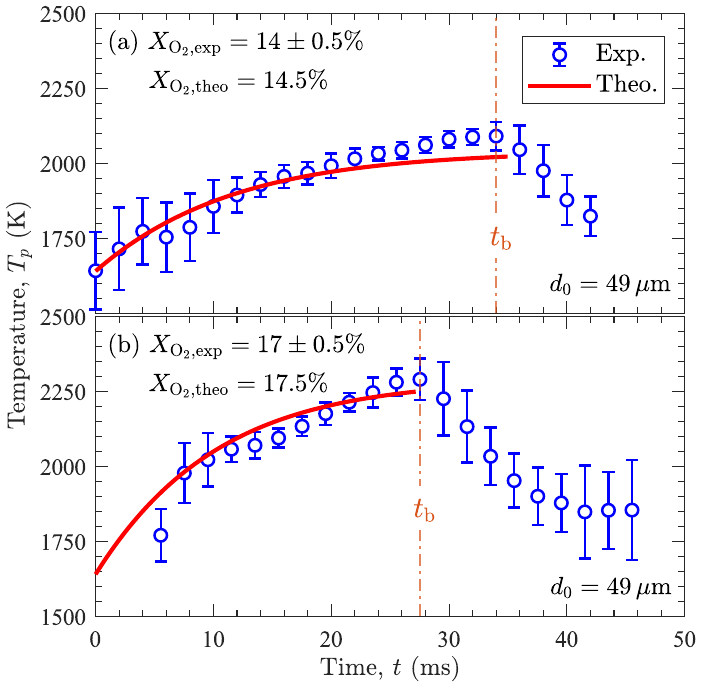}
\caption{Theoretical and experimental temperature histories of $49\,$\textmu m iron particles burning at $14\pm0.5$\% and $17\pm0.5$\% oxygen mole fractions, respectively. The solid curves represent theoretically calculated results for $49\,$\textmu m particles without radiative heat loss. The symbols denote experimentally obtained mean results with standard deviation for $49\pm 5\,$\textmu m particles, as reported in Ref. \cite{ning2023size}. For the sake of clarity, the experimental data are plotted in every \SI{2}{ms}.}
\label{fig: Particle temp history}
\end{figure}
Finally, the weak coupling between burn time and particle temperature suggests that while burn time is a necessary parameter, it alone is not sufficient to validate particle combustion modes. Time-resolved particle temperature emerges as a crucial experimental information for validating theoretical and numerical models of particle combustion.  \Cref{fig: Particle temp history} shows a comparison between the time histories calculated using the current theory and those measured using two-color pyrometry \cite{ning2023size}. 
\begin{comment}
    In contrast to previous experiments \cite{ning2021burn,ning2022temperature}, a higher laser power was used to ignite the particles, resulting in an initial heating of the particles to a relatively high temperature, which was measured directly \textcolor{black}{at a relatively low oxygen level (i.e., 14\%). As the oxygen level increases, the span between the particle peak temperature and intial temperature exceeds the the dynamic range of the two-color pyrometer. In the experiment, the initial temperature was not covered.}
\end{comment}
\textcolor{black}{For the theoretical calculation, the initial temperature and mean particle size are taken from the experiment.  It worth noting that the initial particle temperature was only directly measured at the relatively low oxygen mole fraction of 14\%. At the oxygen mole fraction of 17\%, the same initial particle temperature is assumed in the theory because the particles were heated up with an identical laser power of \SI{40}{W}.}
As illustrated in \cref{fig: Particle temp history}(a), the time history of the theoretical particle temperature generally aligns well with the measurement. Approaching the maximum particle temperature, the theory slightly underpredicts the measurement. The comparison ends at the peak temperature, since further oxidation beyond stoichiometric \ce{FeO} during subsequent reactive cooling is not considered.  As at higher oxygen mole fraction of 17$\pm$0.5\%, where the particle temperature right after ignition was not measured, the initial event of the measured temperature  trajectory cannot represent the start of the combustion process. To establish an absolute time coordinate for the comparison with the theory, in \cref{fig: Particle temp history}(b) the measured temperature profile is shifted  to match with the burn time extrapolated from a fitted correlation between burn time and the initial particle diameter \cite{ning2024experimental}. 
The good agreement confirms that the current theoretical model accurately predicts the time-dependent temperature of burning iron particles, providing a robust validation for scenarios where the particle peak time is much lower than the boiling point of iron, leading to a negligible impact of evaporation.

\begin{comment}
    the oxygen levels in the ambient are not above the air condition.
\end{comment}

%\subsection{Sensitivity analysis}

\section{Conclusions and outlook}
\label{sec: Conclusions}
This work introduces a quantitative analytical theory for the diffusion-limited combustion of nonvolatile (metal) particles. 
Analytical solutions for the burn time and time-dependent particle temperature are derived from the conservation equations. A simple and effective approach is proposed to couple the solutions, which demonstrates a remarkable predictive capability without relying on numerical or experimental inputs. The application of the model to analyze iron particle combustion provides valuable insights into the underlying processes.

First of all, the peak temperature of the particle is constrained by the burnout of iron rather than the balance between the heat generation and heat loss of the particle. When radiation is excluded, the time-dependent parameters of the particle, including temperature, diameter, and the fraction of unburnt metal, follows a scaling of $1/d_0^2$ along the time coordinate. Taking the surface radiation into account do not change the scaling significantly, whereas it introduces a negative size-dependence of the particle temperature. 

Furthermore, the gas-phase pressure, if not very low, plays no role in the diffusion-limited combustion process of nonvolatile particles when the evaporation is negligible. In contrast, increasing the far-field gas temperature accelerates the combustion process, which results in an elevated particle peak temperature following a nearly linear trend with respect to the gas temperature. As the the mass diffusivity of the oxygen is insensitive to the initial particle temperature, the burn time in the diffusion limited regime only slightly changes even for large variations of the initial particle temperature. By the time when the particle peak temperature is reached,  the initial temperature has been largely relaxed. Consequently, the peak temperature is weakly influence by the initial condition.

Finally, the theoretical predictions for the burn time and temperature of isolated iron particle are compared with their experimental counterparts.  
The  quantitatively accurate alignment between the theoretical and experimental results validates the model within the scope of its applicable assumptions.
The presented theory not only stands as a valuable tool for predicting combustion characteristics of iron particles but also serves as a foundation for future studies exploring the intricacies of nonvolatile particle combustion in diverse scenarios.

\section*{Declaration of competing interest}
The authors declare that they have no known competing financial interests or personal relationships that could have appeared to
influence the work reported in this paper.

\section*{Acknowledgments}
This work is founded by the Hessian Ministry of Higher Education, Research, Science and the Arts under the cluster project Clean Circles. 


%% References


\bibliographystyle{model1-num-names}
\bibliography{sample.bib}






% \end{thebibliography}

\newpage
\appendix
\section{Size-dependent particle density }
\label{App:A}
An universal expression of the oxidation reaction of metal with oxygen reads
\begin{equation}
    \mathrm{O_2}\,\mathrm{(g)} + \frac{2x}{y}\mathrm{M}\,\mathrm{(c)} = \frac{2}{y}\mathrm{M}_x\mathrm{O}_y\,\mathrm{(c)}.
    \label{eq:emc}
\end{equation}
Therefore, a stoichiometric oxide-to-oxygen mass ratio  can be defined as
\begin{equation}
    s:=\frac{2}{y}\frac{\mathcal{M}\mathrm{_{mo}}}{\mathcal{M}\mathrm{_\mathrm{O_2}}},
    \label{eqn: stiochiometric ratio}
\end{equation}
with $\mathcal{M}_\mathrm{mo}$ and $\mathcal{M}_{\mathrm{O_2}}$ the molar mass of the metal oxide (M$_x$O$_y$) and that of O$_2$.
The mass of the particle encompassing those of total oxygen and metal elements reads
\begin{equation}
\label{eq:particle mass}
   m_\mathrm{p} = m_\mathrm{M} + m_\mathrm{O},
\end{equation}
where subscripts, M and O, denote the metal and oxygen elements, respectively. The total mass of metal elements is essentially the initial mass of the metal particle:
\begin{equation}
   m_\mathrm{M} = m_0 = \frac{\pi}{6}\rho_\mathrm{m}d_0^3,
   \label{initial mass}
\end{equation}
with d$_0$ the initial particle diameter and $\rho_\mathrm{m}$ the density of the metal. Because of the conservation of O elements, the total mass of O elements in the particle equals that of consumed O$_2$:
\begin{equation}
   m_\mathrm{O} = m_\mathrm{O_2}.
   \label{eqn: oxygen mass}
\end{equation}
According to \cref{eqn: stiochiometric ratio,eq:particle mass,eqn: oxygen mass}, the mass of metal oxide is:
\begin{equation}
   m_\text{mo} = sm_\mathrm{O_2} = sm_\mathrm{O}=  s(m_\mathrm{p}-m_\mathrm{M}) =  s(m_\mathrm{p}-m_0),
\label{mass oxide}
\end{equation}
The remaining mass of the metal is
\begin{equation}
   m_\text{m} = m_\text{p} - m_\text{mo} = (1-s)m_\text{p} + sm_0.
\label{mass metal}
\end{equation}
Assuming the volumetric shrinkage due to mixing between liquid metal and liquid metal oxide (if they are miscible) is negligible, the volume of the particle is
\begin{equation}
   \frac{m_\mathrm{p}}{\rho_\mathrm{p}} = V_\mathrm{p} = V_\text{mo} + V_\text{m} = \frac{m_\text{mo}}{\rho_\text{mo}} + \frac{m_\text{m}}{\rho_\text{m}}.
   \label{particle volume}
\end{equation}
Substituting \cref{mass oxide,mass metal} into \cref{particle volume} yields

\begin{equation}
\label{particle density}
\begin{split}
   \rho_\mathrm{p} &= \frac{\rho_\mathrm{m}\rho_\mathrm{mo}}{s\rho_\mathrm{m}+(1-s)\rho_\mathrm{mo}} + \frac{(\rho_\mathrm{m}-\rho_\mathrm{mo})s\rho_\mathrm{m}}{s\rho_\mathrm{m}+(1-s)\rho_\mathrm{mo}}\frac{V_0}{V_\text{p}}\\
   &= \frac{\rho_\mathrm{mo}}{s+(1-s)\rho_\mathrm{mo}/\rho_\mathrm{m}} + \frac{(\rho_\mathrm{m}-\rho_\mathrm{mo})s}{s+(1-s)\rho_\mathrm{mo}/\rho_\mathrm{m}}\frac{V_0}{V_\text{p}}
\end{split}   
\end{equation}
Now let's define a dimensionless parameter, oxide-to-metal density ratio:
\begin{equation}
\label{density ratio}
   \varrho := \frac{\rho_\text{mo}}{\rho_\text{m}}.
\end{equation}
Introducing another dimensionless parameter, diameter expansion ratio of the particle:
\begin{equation}
\label{expansion ratio}
   \epsilon := \frac{d_1}{d_0} = \sqrt[3]{\frac{m_1/\rho_\text{mo}}{m_0/\rho_\mathrm{m}}} = \sqrt[3]{\frac{\mathcal{M}_\text{mo}/x}{\mathcal{M}_\text{m}}\frac{1}{\varrho}},
\end{equation}
with $d_1$ the particle diameter at burnout, $\mathcal{M}_\text{m}$ the molar mass of the metal, $\varrho$ also reads
\begin{equation}
\label{density ratio with particle diameter}
   \varrho = \frac{1}{\epsilon^3}\cdot\frac{\mathcal{M}_\text{mo}}{x\mathcal{M}_\text{m}} = \left(\frac{d_0}{d_1}\right)^3\frac{\mathcal{M}_\text{mo}}{x\mathcal{M}_\text{m}}.
\end{equation}
\Cref{density ratio with particle diameter} allows us to determine $\varrho$ by measuring the diameter expansion ration, $\epsilon$. Therefore, estimating the densities of liquid metals and liquid oxides becomes unnecessary if they are not known \textit{a priori}. However, the oxidation state at burnout is still needed, which may be clear for some metals, such as FeO for iron at approximately the end of diffusion-limited combustion.
Then substituting \cref{density ratio} into \cref{particle density} gives:
\begin{equation}
\label{particle density with sigma}
   \rho_\text{p} = \frac{\rho_\text{mo}}{s+(1-s)\varrho} + \frac{(\rho_\text{m}-\rho_\text{mo})s}{s+(1-s)\varrho}\frac{V_0}{V_\text{p}}
\end{equation}
Assuming that the particle remains spherical during oxidation, the particle volume then expresses as:
\begin{equation}
   v_\mathrm{p}=\frac{\pi}{6}d_\mathrm{p}^3.
   \label{particle volime}
\end{equation}
Therefore, the particle density as a function of diameter reads
\begin{equation}
\label{particle density with diameter}
\begin{split}
   \rho_\mathrm{p} &= \frac{\rho_\mathrm{mo}}{s+(1-s)\varrho} + \frac{(\rho_\mathrm{m}-\rho_\mathrm{mo})s}{s+(1-s)\varrho}\left(\frac{d_0}{d_\mathrm{p}}\right)^3 \\
   &=\rho_1 + \rho_2\left(\frac{d_0}{d_\mathrm{p}}\right)^3,
\end{split}
\end{equation}
with
\begin{equation}
    \rho_1 := \frac{\rho_\mathrm{mo}}{s+(1-s)\varrho},
\end{equation}
and
\begin{equation}
    \rho_2 := \frac{(\rho_\mathrm{m}-\rho_\mathrm{mo})s}{s+(1-s)\varrho}.
\end{equation} 
Although \cref{particle density with diameter} has been obtained previously by Hazenberg \cite{hazenberg2019eulerian}, the alternative derivation procedure elaborated here is new and more straightforward.

\section{Composition of multi-species gas at the film layer}
\label{App:B}
 When only two gaseous species (i.e., oxygen and inert gas) exist, the gas composition in the film layer can be easily determined using \cref{averaging rule}. When several inert species present, the determination of the gas composition in the film is a bit complicated (we only consider the situation where only one kind of oxidizer exists). To accomplish this task, we first define a mean molar mass of inert species:
\begin{equation}
    \overline{\mathcal{M}_\mathrm{I}} := \frac{\sum_{j \neq j_\mathrm{O}}^{N}{X_j \mathcal{M}_j}}{\sum_{j \neq j_\mathrm{O}}^{N}{X_j}},
\end{equation}
where $N$ is the number of species and $j_\mathrm{O}$ is the index of the oxygen. As the molar ratios between inert species are constant, $\overline{\mathcal{M}_\mathrm{I}}$ is constant too, provided that the diffusivities of all the inert species are identical. At the film layer, the mass fraction of the oxygen can be formulated with $\overline{\mathcal{M}_\mathrm{I}}$:
\begin{equation}
    Y_\mathrm{O,f} = \frac{X_\mathrm{O,f}\mathcal{M}_\mathrm{O}}{X_\mathrm{O,f}\mathcal{M}_\mathrm{O} + (1-X_\mathrm{O,f})\overline{\mathcal{M}_\mathrm{I}}}.
\label{eqn: oxygen mass fraction in film}
\end{equation}
Rearranging \cref{eqn: oxygen mass fraction in film} yields the molar fraction of the oxygen at the film layer:
\begin{equation}
    X_\mathrm{O,f} = \frac{Y_\mathrm{O,f}\overline{\mathcal{M}_\mathrm{I}}}{Y_\mathrm{O,f}\overline{\mathcal{M}_\mathrm{I}}+(1-Y_\mathrm{O,f})\mathcal{M}_\mathrm{O}}.
\end{equation}
The molar fractions of inert species are than can be determined as
\begin{equation}
    X_{j,\mathrm{f}} = (1-X_\mathrm{O,f})\frac{X_{j,\infty}}{\sum_{j \neq j_\mathrm{O}}^{N}X_{j,\infty}},\ j\in [1,N]\ \&\ j\neq j_\mathrm{O}.
\end{equation}
Finally, $\lambda_\mathrm{f} =\lambda(T_\mathrm{f}, X_\mathrm{f})$, $(\rho_\mathrm{g} \mathcal{D}_\mathrm{O})_\mathrm{f} = \rho_\mathrm{g}(T_\mathrm{f}, X_\mathrm{f}) \mathcal{D}_\mathrm{O}(T_\mathrm{f}, X_\mathrm{f})$, $c_{p,\mathrm{f}} = c_{p,\mathrm{g}}(T_\mathrm{f}, X_\mathrm{f})$, and $c_{p,\mathrm{O}} = c_{p,\mathrm{O}}(T_\mathrm{f})$ are determined. $X_\mathrm{f}$ is the array of the molar fractions of all gaseous species at the film layer.

\end{document}